\documentclass[aip,pof]{revtex4-1}
\usepackage{graphicx}
\usepackage{color}
\usepackage{amsmath}
\usepackage{amssymb}
\usepackage{dcolumn}
\usepackage{bm}
\usepackage{tabularx}
\usepackage{longtable}
\definecolor{grey0}{gray}{0.8}
\definecolor{grey1}{gray}{0.75}
\definecolor{grey2}{gray}{0.5}
\definecolor{grey3}{gray}{0.35}

\begin{document}

\title{A minimal coupled fluid-discrete element model for bedload transport}
\author{R. Maurin}
\email{raphael.maurin@irstea.fr}
\affiliation{Irstea, Grenoble, UR ETGR, 2 rue de la papeterie, BP 76, 38402 Saint-Martin-d'H\`eres, France\\ Univ. Grenoble Alpes, Irstea, F-38000 Grenoble, France}
\author{J. Chauchat}
\affiliation{Univ. Grenoble Alpes, LEGI, F-38000 Grenoble, France\\
CNRS, UMR 5519, LEGI, F-38000 Grenoble, France}
\author{B. Chareyre}
\affiliation{Univ. Grenoble Alpes, 3SR, F-38000 Grenoble, France\\
CNRS, UMR 5521, 3SR, F-38000 Grenoble, France}
\author{P. Frey}
\affiliation{Irstea, Grenoble, UR ETGR, 2 rue de la papeterie, BP 76, 38402 Saint-Martin-d'H\`eres, France\\ Univ. Grenoble Alpes, Irstea, F-38000 Grenoble, France}

\begin{abstract}
A minimal Lagragian two-phase model to study turbulent bedload transport focusing on the granular phase is presented, and validated with experiments. The model intends to describe bedload transport of massive particles in fully rough flows at relatively low Shields numbers, for which no suspension occurs. A discrete element method for the granular phase is coupled with a one dimensional volume-averaged two-phase momentum equation for the fluid phase. The coupling between the discrete granular phase and the continuous fluid phase is discussed, and a consistent averaging formulation adapted to bedload transport is introduced. An original simple discrete random walk model is proposed to account for the fluid velocity fluctuations. The model is compared with experiments considering both classical sediment transport rate as a function of the Shields number, and depth profiles of solid velocity, volume fraction, and transport rate density, from existing bedload transport experiments in inclined flume. The results successfully reproduce the classical 3/2 power law, and more importantly describe well the depth profiles of the granular phase, showing that the model is able to reproduce the particle scale mechanisms. From a sensitivity analysis, it is shown that the fluctuation model  allows to reproduce a realistic critical Shields number, and that the influence of the granular parameters on the macroscopic results are weak. Nevertheless, the analysis of the corresponding depth profiles reveals an evolution of the depth structure of the granular phase with varying restitution and friction coefficients, which denotes the non-trivial underlying physical mechanisms. 
\end{abstract}


\maketitle

\section{Introduction}
\label{intro}

Historically studied by hydraulic engineers in relation to the management of river waterways \cite{DuBoys1879}, bedload represents the main contribution of sediment transport to the evolution of riverbeds. As such, it has major implications for environmental flows and associated risks like floods for example. In contrast to suspension, bedload transport is characterized by sediment transport for which the vertical gravity force is on average stronger than the upward fluid force, i.e. sediments rolling, sliding or in saltation over the bed. The paper focuses on bedload transport in turbulent flow conditions, which is the most common case in nature.\\

In this phenomenon, one of the main challenges is to link the sediment transport rate to the fluid flow rate. By making the problem dimensionless, it is equivalent to linking the dimensionless sediment transport rate $Q_s^* = \frac{Q_s}{d\sqrt{(\rho^p/\rho^f -1)gd}} $, to the Shields number which compares the fluid bed shear stress $\tau^f_b$ to the buoyant weight $\theta = \frac{\tau^f_b}{(\rho^p-\rho^f) g d}$; where $Q_s$ is the sediment transport rate per unit width, $d$ is the particle diameter, $\rho^p$ and $\rho^f$ are respectively the particle and fluid density, and g is the acceleration of gravity. The usual semi-empirical formulas established in this framework such as the Meyer-Peter and M\"uller \cite{MPM1948} one, can differ by two orders of magnitude from what is observed in the field \cite{Bathurst2007}. This difference is usually explained by the difficulty of measurements, the complexity of the physical processes and the great variability of the situations encountered in the field (e.g. grain size segregation, particle shape, channel geometry)\cite{Recking2010}. \\

Bedload transport can be viewed as a granular medium in interaction with a fluid flow. Following this two-phase decomposition, there are two major possibilities for numerical modelling: a continuous description for the two phases (Euler/Euler) or a continuous description for the fluid phase and a discrete one for the granular phase (Euler/Lagrange). The former considers the momentum conservation of the two phases viewed as two continua in interaction, and is based on the two-phase Reynolds Averaged Navier-Stokes (RANS) equations \cite{Jackson2000,Drew1999}. The averaged equations require different closures, and the main differences between the models pertain to the Reynolds stress tensor and the constitutive law for the intergranular stress. The Reynolds stress tensor models the effect of turbulence on the mean fluid flow, and ranges from simple descriptions such as mixing length formulations, to more complex ones such as $k-\epsilon$. In the case of intense bedload transport, also termed sheet flow, a substantial number of particle layers are in motion. The Euler/Euler description has therefore been mainly used for this regime, with closures for the granular stress tensor according to the main theories for granular media, i.e. Bagnold formulation \cite{Hanes1985}, the $\mu(I)$ rheology \cite{RevilBaudard2013,Aussillous2013}, or the kinetic theory \cite{Jenkins1998,Hsu2004}. \\

The continuous approximation breaks down for the granular phase when considering bedload transport closer to the threshold of motion which is the common situation in mountain streams and the focus of this paper. Moreover, the constitutive equation for granular media is still a matter of debate and thus limits the analysis of the results of such models. Euler/Lagrange models overcome these two limitations by resolving the motion of each grain. For bedload transport, the high concentration of particles inside the bed requires modelling the contact between grains, this is today commonly handled with the Discrete Element Method (DEM). The different scales of fluid description range from large scale average description, to Direct Numerical Simulation (DNS) resolving the fluid locally around the particles down to the smallest turbulence length scale. Euler/Lagrange approaches have been intensively developed in recent years for problems with particles in fluids such as fluidized bed \cite{Tsuji1993,Zhu2007}, particle suspension \cite{Kidanemariam2013}, or sheet flow \cite{Jiang1993,Yeganeh2000,Drake2001,Calantoni2006,Yeganeh-Bakhtiary2009,Kidanemariam2014}. Focusing on bedload transport, up to now only a few contributions have taken advantage of the Eulerian/Lagrangian approach. The work of Duran {\it et al.} \cite{Duran2012} used a simple average description of the fluid with a two dimensional discrete element method for the particles, to numerically study the dependence of sediment transport on the solid-fluid density ratio. Bedload transport was considered in this paper as an extreme case of low density ratio, the closures of the model being more adapted to aeolian transport. Using a DNS/DEM model, Ji {\it et al.} \cite{Ji2013} focused on the influence of particle transport on the turbulence. With a similar model, Fukuoka {\it et al.} \cite{Fukuoka2014} studied particle shape influence and size-segregation effects. \\

Bedload transport has mainly been studied focusing on the fluid phase. It is however clear that the granular behavior is important in this phenomenon and should be studied further \cite{Frey2009,Frey2010}. The idea is therefore to analyze bedload transport at the particle scale in order to understand the behavior of the bed as a granular medium. Considering the complexity of the experimental technique for particle-scale three dimensional (3D) bedload transport analysis (e.g. index matching \cite{Aussillous2013}, or Magnetic Resonance Imaging \cite{Kawaguchi2010}), there are interests in developing a numerical approach of the problem. Focusing on the granular phase, the paper presents a model for bedload transport using a DEM Lagrangian description of the granular phase coupled with a one dimensional volume-averaged two-phase momentum equation for the fluid phase. Although this type of model is common, to our knowledge, no previous contribution focused on bedload transport at relatively low Shields number. Moreover, the usual experimental validations are limited to the classical macroscopic results of dimensionless transport rate as a function of the Shields number. In this paper, we present a model adapted to subaqueous bedload transport (section \ref{Model}) and perform a new particle-scale experimental comparison with solid depth profiles in quasi-2D bedload transport cases \cite{Frey2014} (section \ref{expComparison}). In addition, the classical experimental comparison of the sediment transport rate as a function of the Shields number is considered in a more general 3D framework (section \ref{Analysis}). The influence of the different model contributions is considered in terms of sediment transport rate and solid depth profiles.

\section{Numerical model formulation}
\label{Model}

The proposed model is based on a DEM Lagrangian description for the solid phase and an Eulerian description for the fluid phase. In the present approach, the fluid flow is not solved at the particle scale and the momentum coupling is ensured in an averaged sense via semi-empirical correlations. After presenting briefly the Discrete Element Method (section \ref{solid}) and the fluid phase description (section \ref{fluid}), the coupling between both phases is discussed by detailing in particular the averaging procedure and the coupling forces in the framework of bedload transport (section \ref{coupling}).

\subsection{Solid phase}
\label{solid}

The DEM, originally introduced by Cundall \& Strack \cite{Cundall1979} for granular media, is based on the explicit resolution of Newton's equation of motion for each individual particle considering nearest neighbor contact forces $\vec{f_c^p}$. For each particle $p$ at position $\vec{x}^p$ the equation of motion reads: 
\begin{equation}
m \frac{d^2 \vec{x}^p}{d t^2} = \vec{f}_c^p + \vec{f}_g^p + \vec{f}_f^p,
\label{DEM}
\end{equation}
where $\vec{f_g^p}$ is the force due to gravity and $\vec{f}_f^p$ represents the forces applied by the fluid on particle $p$. This last term arises from the DEM-fluid coupling and will be detailed in subsection \ref{coupling}. The application of the gravity force is straightforward. The contact forces are determined from the relative displacement of the neighboring particles using a defined contact law. In bedload transport, there is a sharp transition between rapidly sheared particles at the interface with the fluid, and almost quasi-static motion inside the bed. The so-called spring-dashpot contact law used in this paper, allows description of these two types of behavior and is classical in granular flows modelling. The contact law is based on a spring of stiffness $k_n$ in parallel with a viscous damper of coefficient $c_n$ for the normal contact, coupled with a spring of stiffness $k_s$ associated with a slider of friction coefficient $\mu$ for the tangential contact. For normal contact, the linear elastic spring and viscous damping define a constant restitution coefficient $e_n$ characteristic of the energy loss at collision, which can be evaluated experimentally.

\subsection{Fluid phase}
\label{fluid}
\begin{figure}
\centering
\includegraphics[width=0.6\textwidth]{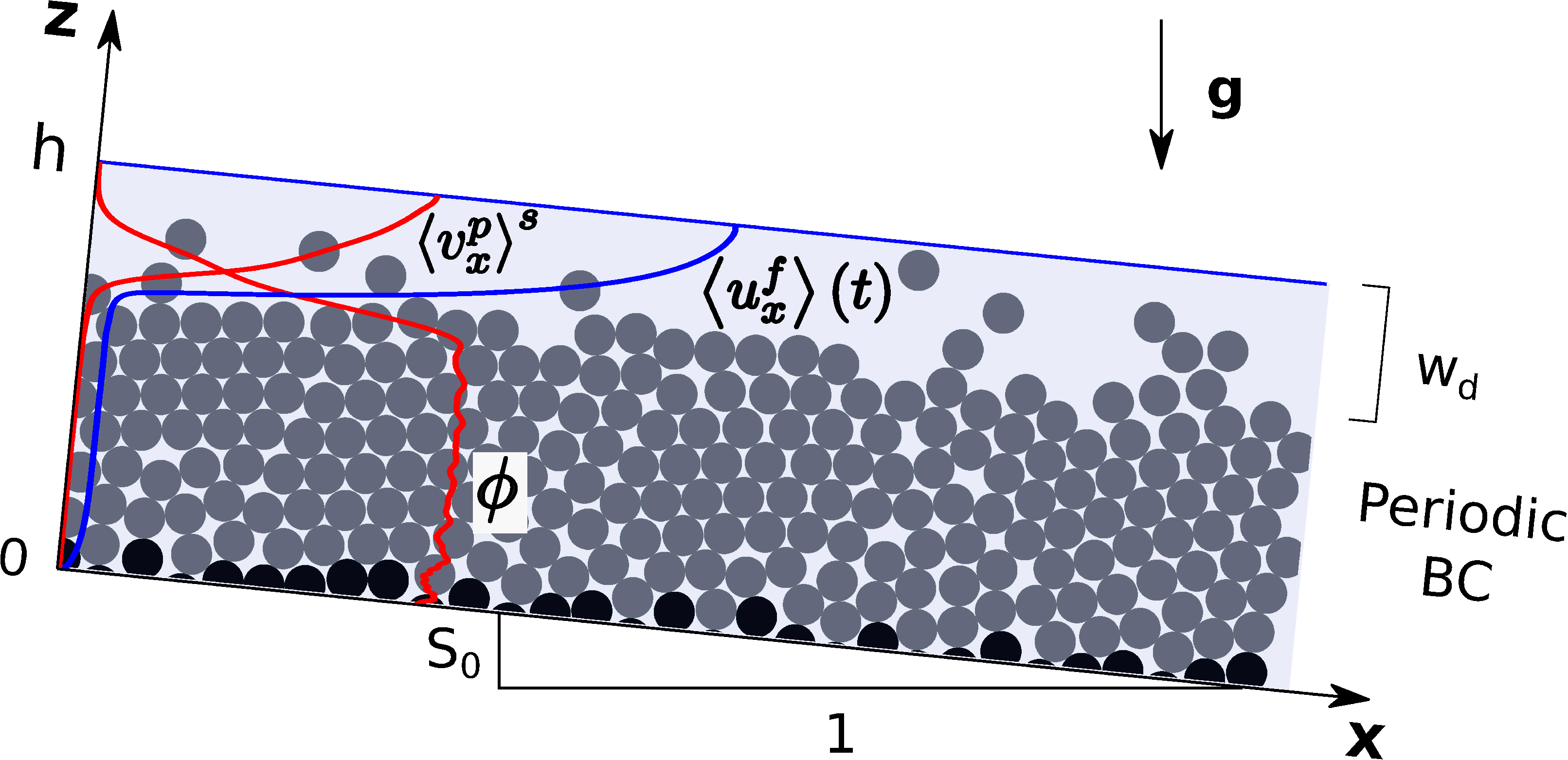}
\caption{\label{situationScheme}Sketch of the problem representing the axes and the variables used in the model: the water free-surface position $h$, the water depth $w_d$, the slope $S_0 = \tan \alpha$, the gravity vector $\vec{g}$, as well as the profiles of average streamwise fluid and solid velocities (resp. $\left<u_x\right>^f$ and $\left<v_x^p\right>^s$), and solid volume fraction $\phi$. Streamwise periodic boundary conditions (BC) are used for the solid phase DEM description as indicated on the scheme.}
\end{figure}

The fluid phase model is based on spatially averaged two-phase Navier-Stokes equations \cite{Jackson2000}, and inspired from the one-dimensional Euler-Euler model proposed by Revil-Baudard \& Chauchat \cite{RevilBaudard2013} to deal with turbulent unidirectional sheet-flows. The simplifications of the general fluid phase equations \cite{Jackson2000} due to the incompressible and unidirectional character of the present bulk flow lead to the resolution of the same fluid phase momentum equation:
\begin{equation}
 \epsilon \ \rho^f \frac{\partial \left<u_x\right>^f}{\partial t} =  \frac{d~(\epsilon\left<\tau_{xz}\right>^f)}{dz} + \frac{d R_{xz}^f}{dz} + \epsilon \rho^f g \ \sin \alpha - n \left<f_x\right>^s,
\label{momentumBalance}
\end{equation}
where $\epsilon$ is the fluid phase volume fraction, $\rho^f$ is the fluid density, $ \left< u_x \right>^f$ is the averaged fluid velocity, $\left<\tau_{xz}\right>^f$ is the averaged fluid viscous shear stress, $R_{xz}^f$ is the Reynolds shear stress, $\alpha$ is the channel inclination angle, and $n\left<f_x \right>^s$ is the averaged fluid-particle general interaction term. A schematic picture representing the main fluid model variables is shown in figure \ref{situationScheme}.\\

The operator $\left< .\right>^s$ denotes a spatial averaging over the solid phase while the operator $\left< .\right>^f$ denotes a spatial averaging over the fluid phase. The major difference with the continuous two-phase model proposed by Revil-Baudard \& Chauchat \cite{RevilBaudard2013} and with Euler/Euler models in general, stands in the average fluid-particle interaction $n\left<f_x \right>^s$ and solid volume fraction $\phi = 1- \epsilon$, obtained in the present model from a spatial averaging of the DEM solution, whereas otherwise obtained by solving a continuous momentum balance equation. All the details concerning the averaging process and the fluid-particle interaction term will be given in subsection \ref{coupling}. \\

In equation (\ref{momentumBalance}), omitting the fluid-particle interaction term, closure laws for the viscous shear stress $\left<\tau_{xz}\right>^f$ and the Reynolds shear stress $R_{xz}^f$ need to be prescribed. In the present model, the fluid is considered as Newtonian, so that:
\begin{equation}
 \left<\tau_{xz}\right>^f = \rho^f \nu^f \frac{d \left<u_x\right>^f}{dz},
\label{viscouStress}
\end{equation}
where $\nu^f$ is the clear fluid kinematic viscosity.\\
The Reynolds shear stress, representing the vertical turbulent mixing of horizontal momentum, is modeled based on the eddy viscosity concept ($\nu^t$) with a mixing length formulation: 
\begin{equation}
 R_{xz}^f =  \rho^f ~ \nu^t \frac{d \left<u_x\right>^f}{dz} \ \ \text{with}\ \ \nu^t = \epsilon \ l_m^2 \left|\frac{d \left< u_x \right>^f}{dz}\right|,
\end{equation}
in which the mixing length $l_m$ formulation proposed by Li \& Sawamoto\cite{Li1995} is used:
\begin{equation}
l_m(z) = \kappa \int_0^z{\frac{\phi^{max} - \phi(\zeta)}{\phi^{max}} ~d\zeta},
\label{mixingLength}
\end{equation}
where $\kappa=0.41$ represents the von Karman constant. This simple formulation allows recovery of the two expected asymptotic behaviors: the mixing length is linear with $z$ when the solid phase volume fraction vanishes (\textit{i.e.} clear fluid), as in the law of the wall \cite{Prandtl1926}, and the mixing length is zero when the solid phase is at its maximum packing fraction, \textit{i.e.} the turbulence is fully damped inside the dense sediment bed. As explained in Revil-Baudard \& Chauchat \cite{RevilBaudard2013}, this formulation is well adapted for boundary layer flow above mobile rough beds. Indeed, the integral of the solid volume fraction predicts a non zero mixing length at the transition between the granular dominated and turbulent dominated layers. Also, with this formulation no virtual origin for the mixing length has to be prescribed.\\

\subsection{DEM-fluid coupling}
\label{coupling}

The key point in continuous-discrete models consists in the coupling of the two phases, which involves an averaging procedure and a parametrization of the fluid forces applied on the particles. 

\subsubsection{Averaging procedure} 
\label{averageProc}
For this purpose, the spatial averaging operator for the solid phase needs to be defined consistently with the spatial averaging operator for the fluid phase \cite{Jackson2000,Anderson1967}. According to Jackson \cite{Jackson2000}, the solid phase volume fraction $\phi(\vec{x_1})$ at a given position $\vec{x_1}$ is defined as:
\begin{equation}
\phi(\vec{x_1}) = \sum_p \int_{V_p} \mathcal{G}(|\vec{x_1}-\vec{x}|) dV
\label{solidFraction}
\end{equation}
where the sum is over all the particles, $V_p$ is the volume of particle $p$, and $\mathcal{G}(\vec{x})$ is a weighting function that must be normalized on the whole physical domain. Providing that the weighting function $\mathcal{G}$ is defined, the solid phase spatial average of a scalar quantity $\gamma$ at a given position $\vec{x_1}$ is defined as:
\begin{equation}
\left< \gamma \right>^s = \frac{1}{\phi(\vec{x_1})}\sum_p \int_{V_p} \gamma(\vec{x}) \mathcal{G}(|\vec{x_1}-\vec{x}|) dV,
\label{averageDef}
\end{equation}

In the general case, Jackson \cite{Jackson2000} uses a radial weighting function $\mathcal{G}$. However in the present case, to match the discretization of the fluid resolution it is more convenient to define a cuboid weighting function. To fulfill the normalization property, a three-dimensional step function is chosen for the weighting function:
\begin{equation}
\displaystyle \mathcal{G}(\vec{x}) = \left\{
    \begin{array}{ll}
\displaystyle    \frac{1}{l_x ~ l_y ~ l_z} & \mbox{for } |x| \leq l_x/2 \mbox{, } |y| \leq l_y/2  \mbox{, } |z| \leq l_z/2 \\[12pt]
        0 & \mbox{otherwise}\\
    \end{array}
\right.
\label{weightingFunction}
\end{equation}

In order to properly define the spatial averaging, the average should be independent from the length scales chosen for the weighting function: $l_x$, $l_y$ and $l_z$ \cite{Jackson2000,Anderson1967}. This is only possible if a separation of scales exists between the macroscopic length scale of the problem $L$, the  length scales associated with the weighting function $l_x,l_y,l_z$ and the particle diameter $d$, \textit{i.e.} $L>>l_x\text{, }l_y\text{, }l_z>>d$. 

Due to the sharp transition occurring at the sediment bed interface in the wall-normal direction, the wall-normal macroscopic length scale of the problem $L$ is lower than the particle diameter $d$. Therefore the vertical length scale of the weighting function $l_z$ should be lower than the particle diameter in order to accurately resolve the vertical gradients of the averaged solid phase variables. We postulate that this limited vertical length scale $l_z$, can be compensated statistically by larger complementary horizontal length scales, $l_x$ and $l_y$. The convergence analysis of the numerical results on the length scale $l_x$ presented in appendix \ref{appendixAverage} lends credibility to this hypothesis.

\subsubsection{Fluid forces}
The force applied by the fluid on a single particle $\vec{f}_f^p$ introduced in equation (\ref{DEM}) is defined as the integral of the total fluid stress, pressure and shear stress, acting on the particle surface \cite{Jackson2000}. In the present model, the fluid flow is not resolved at the particle scale so that the hydrodynamic forces cannot be computed explicitly, and need to be prescribed through semi-empirical formulas based on average fluid variables. The main hydrodynamic forces in bedload transport reduce to the buoyancy, the drag and the shear-induced lift. Ji {\it et al.} \cite{Ji2013} numerical results exhibit a non-negligible importance of the lift force with respect to the other two. However, Schmeeckle {\it et al.}\cite{Schmeeckle2007} showed experimentally that the usual formulation of the lift \cite{Wiberg1985}, derived using the inviscid flow assumption, is not valid close to the threshold of motion. Based on this observation and the absence of alternative formulation, it has been decided not to include the lift force at this stage.\\

Therefore, the force $\vec{f}_f^p$ induced by the fluid on a particle p appearing in the DEM model (equation (\ref{DEM})), reduces to buoyancy $\vec{f}_{b}^p$ and drag $\vec{f}_{D}^p$:
\begin{equation}
\vec{f}_f^p = \vec{f}_{b}^p + \vec{f}_{D}^p.
\label{fluidForce}
\end{equation}
According to Jackson\cite{Jackson2000} the generalized buoyancy force is defined as:
\begin{equation} 
\vec{f}_{b}^p  =  V^p ~\left( -\vec{\nabla}\left<P \right>^f  + \vec{\nabla}.\left<\overline{\overline{{\tau^f}}} \right>\right),
\label{generalisedBuoyancy}
\end{equation}
where $\left<P\right>^f$ is the average fluid pressure and $\left<\overline{\overline{{\tau}}} \right>^f$ is the average viscous shear stress tensor taken at a larger scale than the particle scale. This definition generalizes the so-called Archimedes buoyancy force for hydrostatic problems to cases where the fluid volume is submitted to a macroscopic deformation at a scale much larger than the particle scale \textit{i.e.} the fluid deformation viewed by the particles can be considered as constant. Similarly to Revil-Baudard \& Chauchat \cite{RevilBaudard2013}, we found that the viscous stress tensor contribution is however negligible with respect to the pressure contribution in bedload transport. The force applied on each particle can then be approximated by the usual buoyancy expression, which is equivalent to apply the buoyant weight in the vertical direction.\\

The drag force exerted by the fluid flow on a single particle is classically expressed as:
\begin{equation}
\displaystyle \vec{f}_{D}^p = \frac{1}{2}\rho^f \frac{\pi d^2}{4} ~ C_D ~ \left|\left| \left<\vec{u}\right>^f_{\vec{x^p}} - \vec{v^p} \right|\right|\left(\left<\vec{u}\right>^f_{\vec{x^p}} - \vec{v^p}\right),
\label{drag}
\end{equation}
where $C_D$ is the drag coefficient, and $\left<\vec{u}\right>^f_{\vec{x^p}} - \vec{v^p}$ is the relative velocity between the particle and the average fluid velocity taken at the position of the particle center. The Dallavalle formulation \cite{Dallavalle1948} together with a Richardson-Zaki correction \cite{Richardson1954} is used in the present model for the drag coefficient:
\begin{equation}
C_D = \left(0.4+ \frac{24.4}{Re_p}\right) (1-\phi)^{-\zeta},
\label{CD}
\end{equation}
where $Re_p=|| \left<\vec{u}\right>^f_{\vec{x^p}} - \vec{v^p} || d / \nu^f$ is the particulate Reynolds number for particle $p$. This simple formulation has been used in different two-phase flow models for sediment transport applications \cite{RevilBaudard2013,Jenkins1998,Hsu2004}. The Richardson-Zaki correction $(1-\phi)^{-\zeta}$ accounts for the hindrance effect induced by the local particle concentration, and allows to recover realistic fluid velocity in the particle bed. The exponent has been set to $\zeta = 3.1$ in reference to Jenkins \& Hanes\cite{Jenkins1998}.\\ Equations (\ref{drag}) and (\ref{CD}) are used to compute the drag force on each individual particles in the DEM model  (equation (\ref{DEM})), while the effect of buoyancy is taken into account through the vertical buoyant weight.\\

The average effect of the particles on the fluid momentum balance does not simply consist in the solid averaging of the momentum transfer associated with the hydrodynamic forces. It also includes higher-order correlations which appear in the averaging process, and are due to perturbations of the flow by the presence of the particles. For the case of Stokesian particles at low concentration, Jackson showed analytically \cite{Jackson1997} that these higher-order correlations lead to a modification of the  viscosity in the average viscous fluid stress tensor formulation, which takes the form of Einstein's effective viscosity \cite{Einstein1906}. In the model, the clear fluid viscosity in equation (\ref{viscouStress}) has been replaced by Einstein's effective viscosity $\nu^e$ to take this effect into account:
\begin{equation}
\displaystyle \nu^e = \nu^f \left( 1 + \frac{5}{2} \phi \right).\\
\end{equation}
The phase interaction term in the fluid momentum balance (eq. \ref{momentumBalance}) reduces then to the momentum transfer associated with the hydrodynamic forces. In the present 1D fluid resolution, it is expressed as the average number of particles $n = \phi/V_p = 6 \phi/\pi d^3$ multiplied by the solid-phase average streamwise associated force. For drag force, it gives:
\begin{equation}
n\left<{f_D}_x\right>^s = \frac{3}{4}~\frac{\phi ~ \rho^f}{d} \left<  C_D ~ \left|\left| \left<\vec{u}\right>^f_{\vec{x^p}} - \vec{v^p} \right|\right| ~ \left( \left<u_x\right>^f - v^p_x\right) \right>^s.
\label{averageDrag1final}
\end{equation}
The drag coefficient $C_D$ depends on the relative velocity through the particle Reynolds number, so that it should be included in the spatial averaging. \\

\subsubsection{Velocity fluctuation model}

The proposed average model for the fluid phase does not account for the fluid turbulent velocity fluctuations, which are known in particular to influence the particle threshold of motion. In order to account for these turbulent processes in the average fluid model, a Discrete Random Walk (DRW) model for the fluid velocity fluctuations inspired from Zannetti\cite{Zannetti1986} is therefore introduced. It consists in associating a random velocity fluctuation with each particle for a given duration, as a function of the local turbulent intensity and turbulent time scale. The fluctuations are not correlated in space, nor in time, and the model is built so that the Reynolds shear stress definition is consistent between the average fluid model and the DRW model:
\begin{equation}
\displaystyle \overline{ {u_x^f}' {u_z^f}' } =  - \frac{R_{xz}^f}{\rho^f \epsilon},
\label{fluctProp}
\end{equation}
where the $\overline{\bullet}$ represents an averaging operator in time. 

From experimental measurements in open-channel flows \cite{Nezu1977,Nezu1993}, it has been observed that the magnitude of the fluctuations in the streamwise direction is roughly two times larger than in the vertical direction. With this constraint the following DRW model for the streamwise component $({u_x^f}')^p$ and the normal component $({u_z^f}')^p$ of the fluid velocity fluctuation associated with each particle $p$ is proposed:
\begin{subequations}
 \begin{align}
({u_z^f}')^p &= \lambda_1\\
({u_x^f}')^p &= - ({u_z^f}')^p + \lambda_2,
 \end{align}
\end{subequations}
where $\lambda_1$ and $\lambda_2$ are two Gaussian random numbers of zero mean  and of standard deviation $\sigma$. This standard deviation is obtained from the local value of the Reynolds shear stress at the position of the particle center $\sigma = \sqrt{\frac{R_{xz}^f}{\rho^f \epsilon}(\vec{x^p})}$. The velocity fluctuations are updated every $\tau_t$, defined as the turbulent eddy turn over time, which can be estimated as $\tau_t = w_d/U^f$ where $w_d$ is the water depth, and $U^f$ is the average fluid velocity. These velocity fluctuations are added to the average fluid velocity in the drag force expression (equation (\ref{drag})).\\

\subsection{Numerical resolution strategy}

The resolution of the fluid equation still needs to be clarified. For numerical reasons, it is necessary to express equation (\ref{momentumBalance}) linearly as a function of the average fluid velocity. The numerical treatment of the drag force is then handled as follows:
\begin{equation}
n\left<{f_D}_x\right> = \beta ~ \left( \left<u_x\right>^f - \left<v^p_x\right>^s\right),
\label{averageDrag2}
\end{equation}
where $\beta$ is computed according to equation (\ref{averageDrag1final}) as:
\begin{equation}
\displaystyle \beta =  \frac{3}{4}~\frac{\phi ~ \rho^f}{d} ~ \frac{\left<C_D ~ \left|\left| \left<\vec{u}\right>^f_{\vec{x^p}} - \vec{v^p} \right|\right|\left(\left<u_x\right>^f_{\vec{x^p}} - v^p_x\right)\right>^s}{{\left<u_x\right>^f}_{\vec{x^p}} - \left<v^p_x\right>^s}
\label{beta}
\end{equation}
This formulation allows strictly the same average momentum transfer in the discrete solid phase problem and the continuous fluid phase one. With this definition the fluid phase momentum equation to be solved can be rewritten as: 
\begin{equation}
 \epsilon \rho^f \frac{\partial \left<u_x \right>^f}{\partial t} = \rho^f\frac{\partial}{\partial z} \left[  \left(\epsilon~  \nu^e +  \nu^t\right) \frac{\partial \left<u_x\right>^f}{\partial z} \right]\\
+ \epsilon \rho^f g \ \sin \alpha - \beta \left(\left<u_x\right>^f - \left<v^p_x\right>^s\right)
\label{finalFluidRes}
\end{equation}

This equation is discretized using implicit finite differences for the diffusion and the drag terms. The resulting tridiagonal system is solved using a double-sweep algorithm \cite{Chauchat2013}. The fluid phase resolution period $\tau_f$ should be small enough compared with the particle relaxation time $\tau_{D} = \beta^{-1}$. This characteristic time corresponds to the time needed by a particle initially at rest to reach its steady state velocity in a constant fluid flow. \\

The DEM solid phase model is solved using the open-source code Yade \cite{YADE2010}. The time integration is explicit with a second order centered scheme \cite{Bathe1976} to ensure energy conservation. The time step has been estimated with a method similar to Catalano \cite{Catalano2012} (pp. 84, see also \cite{Catalano2014}), considering the rigidity of the system of springs \cite{Chareyre2005} and dampers as decoupled.

\section{Experimental comparison}
\label{expComparison}

The model is to be compared against experimental data. The declared aim of the present model is to focus on the granular phase behavior. We therefore reproduce the quasi-2D experiments of Frey \cite{Frey2014}, in which particle tracking allowed to obtain average solid depth profiles of bedload transport. After a brief presentation of the experiment, the numerical set-up and the comparison with the experimental results will be presented. 

\subsection{Description of the experiment}
\label{descrExp}
\begin{figure}
  \includegraphics[width=0.5\textwidth]{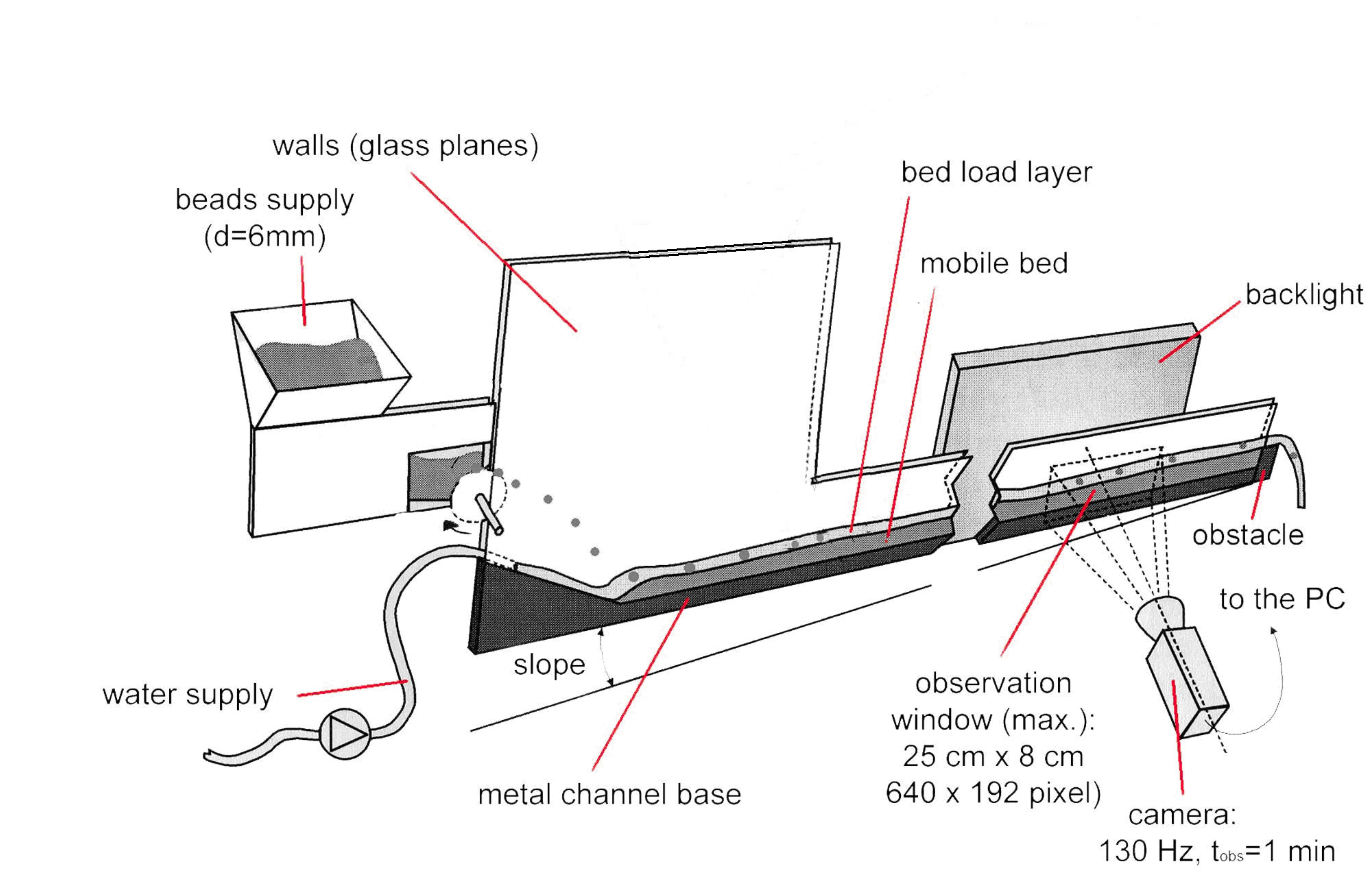}
\caption{\label{expCanal}Experimental setup scheme, modified from B\"ohm {\textit et al}\cite{Bohm2006} and Hergault {\textit et al}\cite{Hergault2010}. The inclined channel width of $6.5d/6$ implies a quasi-2D bead flow, permitting particle tracking of each spherical bead in the observation window filmed by the camera placed perpendicularly.}
\end{figure}

The experiment of Frey \cite{Frey2014} consisted in a quasi-2D ideal case of mountain stream bedload transport on a steep slope. The setup is depicted in figure \ref{expCanal}, it is composed of a $2~m$ long inclined channel of slope $S_0 = 0.1$, and width $6.5d/6$. Water ($\rho^f = 1000~kg/m^3$) flows inside the open-channel and entrains the spherical glass particles ($\rho^p = 2500~kg/m^3$) of diameter $d = 6~mm$. Particles are introduced at the inlet and create an erodible bed thanks to the obstacle placed at the outlet. The number of particle layers is controlled by the height of this obstacle. The channel bottom is made of metal half-cylinders of diameter $d$, fixed at a random elevation between $-2.75~mm$ and $2.75~mm$ with steps of $0.5~mm$ to break clusterization. The particle feeding rate is controlled, and the flow rate is adjusted in order to reach transport equilibrium, i.e. feeding rate equal to the sediment transport rate at the outlet without having aggradation and degradation of the bed. The free-surface fluid flow is turbulent ($Re = U^f w_d/\nu^f \sim 10^4$), hydraulically rough ($Re_p \sim 10^3$), and supercritical ($Fr = U^f/\sqrt{g w_d} \gtrsim 1$). The particle settling velocity ($w_s = 0.54 ~ m/s$) and the suspension number $S^* = w_s/u_*$ are high, meaning that the particles are weakly influenced by the turbulent structures. A camera is placed perpendicular to the sidewall, filming a window of $25\text{x}8~cm^2$ at 131.2 frames per second. Due to the one particle diameter width of the channel, image processing \cite{Bohm2006} enables particle trajectories to be followed inside the measurement window, and the average free-surface elevation to be evaluated. In each experiment, once bedload transport is at equilibrium, data acquisition time lasted $60~s$. Experimental data are averaged in the same way as in the model using the definition of section \ref{coupling}. For more details on the experimental setup, refer to the original experimental article of Frey \cite{Frey2014}. The order of magnitude of the main dimensionless numbers associated with the experiment are shown in table \ref{tableDimensionless}. The Stokes number comparing the inertia of the particle and the viscosity of the fluid is given by $St = \rho^p v^p d/(9\eta^f)$. \\

\begin{table}
\caption{\label{tableDimensionless}Characteristic values of the main dimensionless numbers.}
\begin{ruledtabular}
\begin{tabular}{ccccccc}
$\theta$& $Re$ & $Re_p$ & $Fr$ & $\rho^p/\rho^f$ & $St$ & $S^*$ \\
\noalign{\smallskip}\hline\noalign{\smallskip}
0.05-0.1 & $10^4$ & $10^3$ & $\gtrsim 1$ & 2.5 &  $10^2-10^3$ & $2-10$ \\
\end{tabular}
\end{ruledtabular}
\end{table}

\subsection{Application to the model} 
\label{appModel}

To compare the model with the experiments, the simulation needs to match the experimental set-up. To focus on the bulk equilibrium properties of bedload transport, periodic boundary conditions are considered in the streamwise direction for the present 2D case. The periodic characteristic of the granular phase, does not enable us to impose a feeding rate. To have a situation equivalent to the experiment, the density of beads per unit length (equivalent to the number of layers of particle $N_l$) and the free-surface position $h$ are instead imposed. Indeed, there is a unique couple, slope-water depth, corresponding to the transport equilibrium and it can be reproduced by fixing $h$, $N_l$ and the slope $S_0$ for a periodic sample. In the simulation, for the solid phase, the number of particles is therefore imposed as a function of the length of the periodic cell $l_x$, which has been fixed to $l_x = 1000~d$ to define a consistent and convergent averaging (see appendix \ref{appendixAverage}). The bottom made of fixed particles is randomly generated with the experimental characteristics described in the previous subsection. The boundary conditions for the fluid resolution are imposed considering a fixed boundary at the channel bottom, and forcing the derivative of the fluid velocity to zero at the fixed free-surface elevation measured in the experiment. The other experimental parameters such as the bead size, density and material, or the width of the channel, are set in the simulation at their known experimental values. \\

In the experiment the width to depth ratio is low, and we expect in consequence the fluid flow to have a complex 3D structure. However, experimental flow measurements in this particular channel showed that it still has a typical logarithmic profile \cite{Frey2001}. This, together with the stated aim of the model to focus on the granular phase, made us consider only a correction for the fluid dissipation at the smooth lateral walls. The correction was included as a source term in the fluid averaged momentum balance resolution (eq. \ref{finalFluidRes}), taking the form of a dissipation term evaluated from the classical Einstein method with a Graf and Altinakar friction factor \cite{Graf1998}. The method description can be found in Frey {\textit et al}\cite{Frey2006}. \\

For each run the channel bottom is newly generated randomly, and particles are deposited under gravity. Once the system with fluid resolution is at equilibrium, the simulations last 100 seconds and measurements are made each 0.1 second. The latter corresponds to the particle relaxation time to the fluid velocity $\tau_D = \beta^{-1}$ (eq. \ref{beta}), and is characteristic from the evolution of the system. For the post-processing of both experimental and numerical results, the averaging definition is taken consistently with the numerical resolution from equation (\ref{averagingFormula}).\\

\begin{table}
\caption{\label{tableInput}Model input parameters for the contact law and the fluid resolution. $k_n$ and $k_s$ are respectively the normal and tangential contact stiffness, $e_n$ and $\mu$ denote the restitution and friction coefficient, $\kappa $ is the Von Karman constant, $\zeta$ the Richardson-Zaki exponent, $\phi^{max}$ the bed solid volume fraction, and $\tau_f$ the fluid resolution period.}
\begin{ruledtabular}
\begin{tabular}{cccccccc}
$k_n ~(N/m)$ & $k_s ~(N/m)$ & $e_n$ & $\mu$ & $ \kappa $ & $\zeta$ & $\phi^{max}$ & $\tau_f (s)$\\
\noalign{\smallskip}\hline\noalign{\smallskip}
$5~10^{3}$ & $2.5~10^{3}$  & $0.5$ & $0.4$& $0.41$ & $3.1$ & $0.51$ (2D)/$0.61$ (3D) & $10^{-2}$ \\
\end{tabular}
\end{ruledtabular}
\end{table}

To study the stability of the coupling, we performed a sensitivity analysis on the fluid resolution period, shown in appendix \ref{appendixFluid}, and set it to $\tau_f = 10^{-2}s$ regarding the results obtained. In agreement with Revil-Baudard \& Chauchat\cite{RevilBaudard2013}, it has been found that the fluid effective rheology does not influence the fluid behavior as it is dominated by the turbulent shear stress. We therefore used a clear fluid viscosity. The restitution coefficient was set to $e_n = 0.5$ based on measurements previously made in the experimental channel considered \cite{Bigillon2001}. In the present situation, the limited particle pressure allows artificial softening of the spheres stiffness in order to reduce computational costs. $k_n$ was set to $5.10^3N/m$ which leads to an acceptable average overlap of the order of $10^{-4}d$ and allows to be in the rigid grain limit \cite{Roux2002}. The tangential stiffness was set as a function of the normal one $k_s = k_n/2$. The friction coefficient was taken as the classical value for glass in the dry case $\mu = 0.4$. The main parameters values of the simulation are summarized in table \ref{tableInput}. The simulation results correspond to the application of the experimental conditions, and are not fitted with any parameter afterwards. A summary of the main characteristics of the experimental (Exp) and numerical (Sim) runs is shown in table \ref{tableParam} with respectively the positions of the free-surface $h$ and the number of layers of particle $N_l$ (both measured in the experiments and imposed in the simulation), the measured sediment transport rate expressed in beads per second (b/s) $\dot{n}$, and the measured Shields numbers $\theta$ and $\theta^*$. The latter is defined based on the turbulent shear stress tensor, and can be evaluated only in the simulation, we will come back on the different definition in the next subsection \ref{compResults}. 

\subsection{Results}
\label{compResults}

\begin{figure}
  \includegraphics[width=\textwidth]{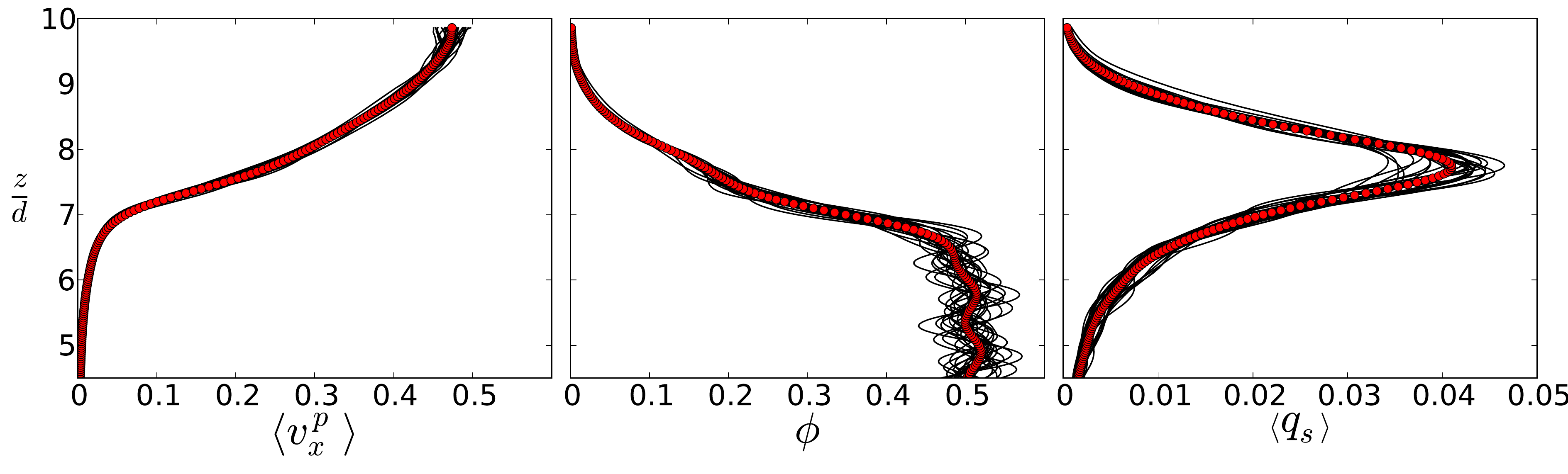}\\
\caption{\label{windowError}Depth profiles of average solid velocity ($m/s$), volume fraction and sediment transport rate density ($m/s$) for the case Sim20 with a periodic length cell of $1000~d$. The different black lines (--) correspond to different post-processing averaging performed in the experimental condition, i.e. over boxes of streamwise length $40~d$ and time-averaged over $60~s$. The full red dots ($\color{red}{\bullet}$) correspond to the averaging as performed for the simulation in general, with a period of averaging of $100~s$ and a streamwise length of the size of the periodic cell. The figure shows the order of magnitude of the variability of the experimental results due to the limited spatio-temporal window of averaging.}
\end{figure}

In bedload transport, one of the main challenges lies in the prediction of the integrated transport rate as a function of the flow rate. The experiment of Frey \cite{Frey2014} was designed to give more insight into the granular behavior, and to focus on the depth profile of bedload transport at the particle scale. It has been noted previously \cite{Frey2014,Lajeunesse2010} that the integrated transport rate per unit width $Q_s$ can be expressed as a function of the average transport rate density $q_s$, the product of the average solid velocity $\left<v_p\right>$ and solid volume fraction $\phi$:
\begin{equation}
Q_s = \int {\left<q_s\right>^s dz} = \int {\left<v^p\right>^s\phi dz}.
\end{equation}
Considering bulk equilibrium properties of bedload transport, $\left<v^p\right>^s$, $\phi$, and $\left<q_s\right>^s$ depends only on the depth $z$. The experimental comparison will then focus on the depth profiles of the solid volume fraction, the average solid velocity, and the transport rate density, which will be called for simplicity transport rate in the following. To complete this decomposition, we evaluated also the Shields number for each case. This was done with two different methods: from macroscopic parameters following Frey \cite{Frey2014}, $\theta = \rho^f Rh_b S_0/[(\rho^p - \rho^f)d]$, with $Rh_b$ the corrected water depth; and from the fluid bottom shear stress defined by the friction velocity $u_*$: $\theta^* = \rho^f u_*^2/[(\rho^p - \rho^f)gd]$, where $u_*$ is given by the maximum turbulent shear stress $u_* = max(R_{xz}^f(z))$. $\theta^*$ was evaluated only in the simulation. This formulation avoids use of the macroscopic determination of the Shields numbers, which is sensitive to the water depth evaluation and the type of wall correction used. \\

\begin{table}
\caption{\label{tableParam}Experimental and numerical run characteristics. The free surface position $h$ and the number of bead layers $N_l$ are both measured in the experiment and imposed in the simulation. $\dot{n}$ is the measured transport rate, $\theta$ and $\theta^*$ the Shields numbers respectively based on macroscopic flow parameters and turbulent shear stress profile. The latter has only been determined in the simulations.}
\begin{ruledtabular}
\begin{tabular}{ccccccc}
Run & h (cm) & $N_l$ & $\dot{n}$ (b/s) & $\theta$ & $\theta^*$ \\
\noalign{\smallskip}\hline\noalign{\smallskip}
Exp6  & 5.3 &  7.08 & 6.67  & 0.076 & - \\
Sim6  & 5.3 &  7.08 & 10.15 & 0.083 & 0.031 \\ 
Exp14 & 5.7 &  7.37 & 13.68  & 0.100 & - \\
Sim14 & 5.7 &  7.37 & 18.13 & 0.120 & 0.048 \\
Exp20 & 5.9 &  7.30 & 19.74  & 0.106 & - \\
Sim20 & 5.9 &  7.30 & 26.38 & 0.130 & 0.061 \\
\end{tabular}
\end{ruledtabular}
\end{table}

In the previous subsections, we did not introduce any experimental or numerical error. It appears that the dispersion is dominated by the limited measurement window length of the experiment ($40~d$). The order of magnitude of this dispersion has been evaluated numerically. Figure \ref{windowError} exhibits the depth profiles of the solid volume fraction, the solid streamwise velocity, and the solid transport rate for a single simulation with different post-processing averaging properties. The fluid mechanics convention is used, where the depth is represented on the y-axis while the quantities of interest are represented on the x-axis. The simulation corresponds to the case Sim20 in table \ref{tableParam}, considering a periodic length cell of $l_x = 1000~d$. The figure shows the variability of the results when the averaging cell length is taken equal to the experimental one at different position in the channel. This dispersion is much greater than the evaluated experimental dispersion and the numerical variability due to the size of the periodic cell simulated. These latter two will be consequently ignored in the comparison, and the variability observed will be taken as error bars. \\
\begin{figure}
  \includegraphics[width=0.95\textwidth]{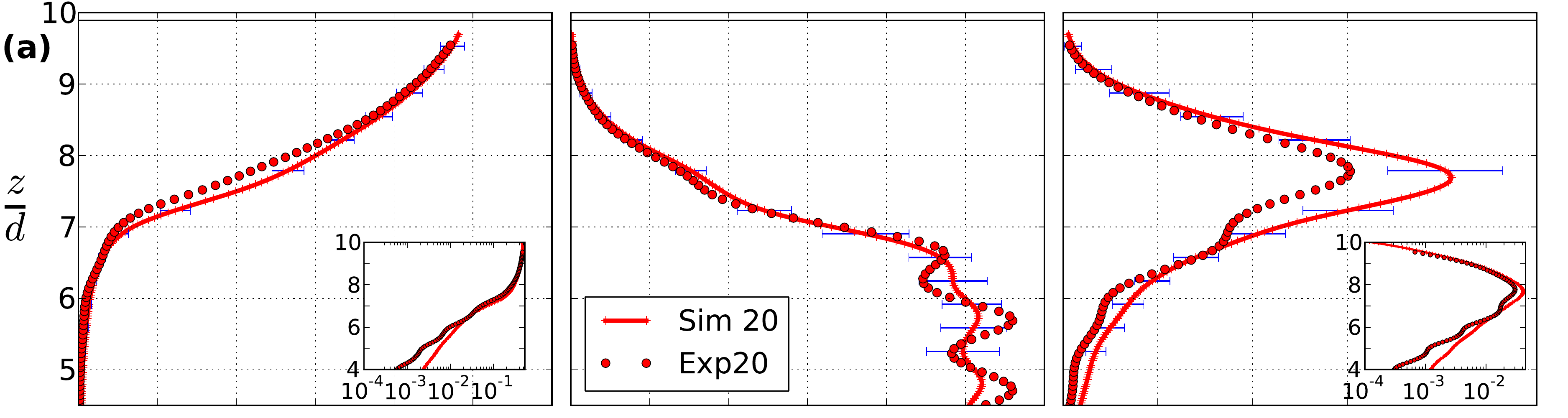}\\
  \includegraphics[width=0.95\textwidth]{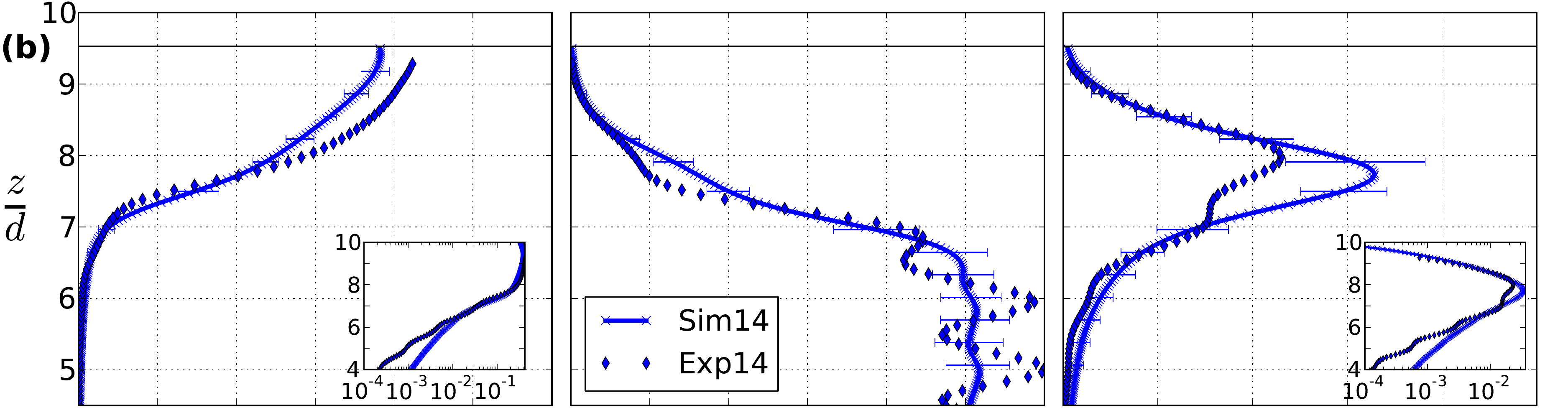}\\
  \includegraphics[width=0.95\textwidth]{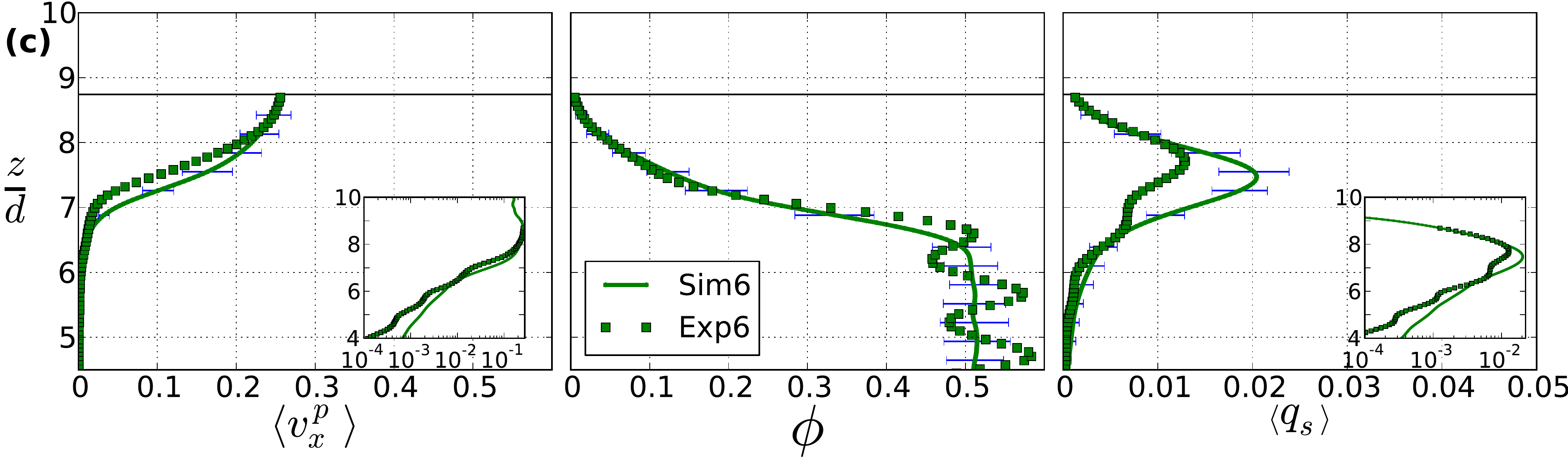}
\caption{\label{resultsFig}Experimental comparison for the different cases presented in table \ref{tableParam}: (a) Case 20, (b) Case 14, (c) Case 6. The figure shows for each case the depth profiles of the average streamwise solid velocity ($m/s$), solid volume fraction, and sediment transport rate density ($m/s$). The full symbol  ($\color{red}{\bullet}$, $\color{blue}{\blacklozenge}$,$\color{green}{\blacksquare}$) represents the experimental results from Frey\cite{Frey2014} while the empty linked one represents the simulation ($\color{red}{+}$, $\color{blue}{\text{x}}$,$\color{green}{\cdot}$). The black line represents the imposed free surface position. The error bars show the variability of the experimental results as evaluated from figure \ref{windowError}. The results show a good general agreement for the three profiles with well-reproduced trends in each case.}
\end{figure}

The three different experiments detailed in table \ref{tableParam} are considered for experimental comparison. The slope is the same and equilibrium transport rate ranges from $6$ to $20~beads/s$. The differences in the input parameters between the runs lie in the water surface position $h$ and the number of layers of particle $N_l$. The different experimental cases represent a good test to evaluate the sensitivity to the parameters and the ability of the model to reproduce different experimental conditions. The macroscopic results presented in table \ref{tableParam} show that the integrated transport rates $\dot{n}$ are in good agreement with the experiment even if slightly overestimated. Considering the Shields number, the two different methods of evaluation lead to an over-estimation using the macroscopic formulation $\theta$, and an under-estimation using the formulation based on the turbulent shear stress $\theta^*$. This underlines the complexity to measure the Shields number in the present case, especially when using the macroscopic formulation which is very sensitive to the small water depth. The trends observed with both formulations are good, and the values have the same order of magnitude than the experiment. In the following, we will use $\theta^*$ in order to avoid the somehow arbitrary determination of the water depth. Using this definition, the value observed for case 6 is below the classical critical Shields number. It should however be kept in mind that the present quasi-2D mono-disperse bed is less resistant, and the value of the critical Shields number is accordingly lowered. To summarize, the general trends observed for the macroscopic parameters are good and these results show that we are able to reproduce the experimental sensitivity to the free surface position and the number of bead layers.\\

\begin{figure}
  \includegraphics[width=\textwidth]{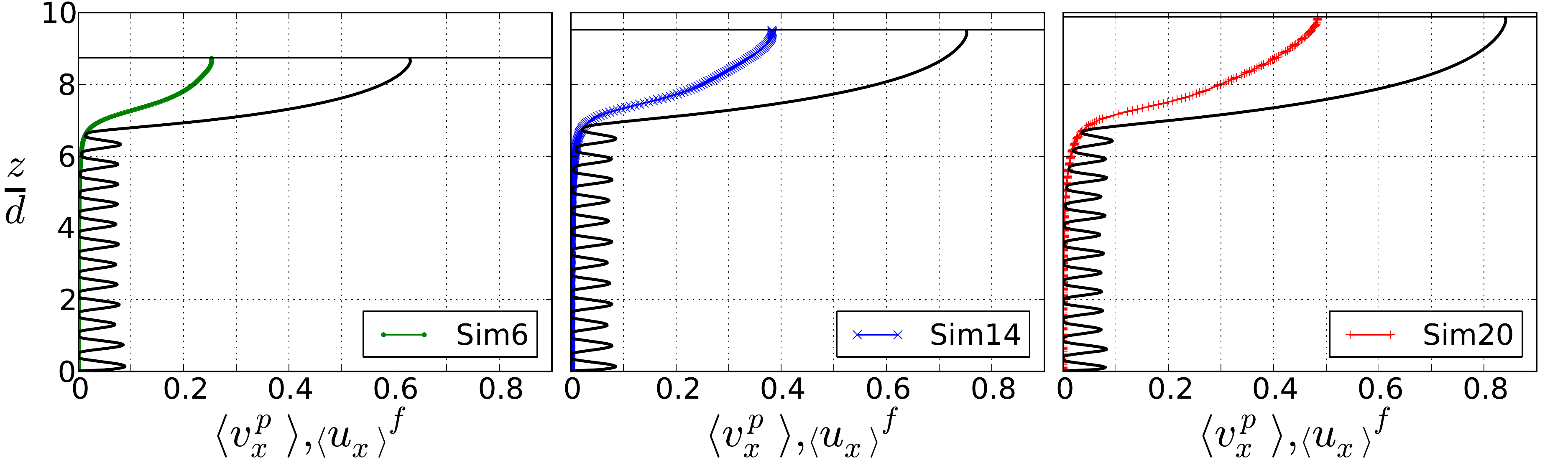}
\caption{\label{fluidProfiles}Simulated solid and fluid (\textbf{--}) velocity profiles for the three cases presented in table \ref{tableParam}.}
\end{figure}

To go further, figure \ref{resultsFig} shows the solid depth profiles of velocity, volume fraction and transport rate density, for the three experimental comparisons. The global trends from one case to the other are well reproduced, and the shape of the simulated curves are close to the experimental ones. Focusing on the transport rate density profile, for each case the value of the peak is slightly overestimated, while the rest of the curve is in very good agreement with experiments. We note an overestimation of the exponential decrease in the bed, in a part weakly affecting the total sediment transport rate density. The oscillations present in each experimental solid volume fraction profile, are representative of the limited size of the experimental averaging window, and impact the sediment transport rate density profile. They are therefore not reproduced in the simulation, and the comparison should be considered with respect to the average value around which it is oscillating. For the solid volume fraction profile, the agreement between simulation and experiments is excellent for case 6 and 20, while we note a slight discrepancy for case 14 at the interface. The solid velocity profiles show a good estimation of the maximal velocity, and of the depth structure. The underestimation of the sediment transport rate peak is shown to correspond to an overestimation of the solid volume fraction in case 14, and of solid velocity in case 6 and 20. \\ For completeness, the fluid velocity profiles are  presented in figure \ref{fluidProfiles}. No experimental data are available for comparison so that the simulated solid velocity profiles have been added for reference. In the fixed bed, the fluid velocity exhibits some oscillations around a finite constant value and the solid velocity is negligible. The oscillations are due to the layering observed in the solid volume fraction profile (figure \ref{resultsFig}) that makes the drag coefficient oscillate accordingly (eq. \ref{CD}-\ref{averageDrag1final}).  In the upper part, the  velocity difference between the solid  and the  fluid phases is  of  the  order of an isolated particle settling velocity $w_s\sim 0.54m/s$. The results are consistent with the drag coefficient formulation adopted and cannot  be  interpreted  further  in the absence of  experimental data.\\

Considering the comparison for the three different cases, with respect to the simplicity of the fluid description and the goal of describing the average solid behavior, the agreement with the experiments is good. The values of the integrated transport rate are close to the experimental ones and the sensitivity to the experimental parameters such as the free-surface position or the number of bead layers has been well reproduced. The comparison of the averaged depth profile of the solid velocity, the solid volume fraction and the transport rate showed that the model is able to reproduce the particle-scale trends observed experimentally, and the variation between the three different runs.

\section{Discussion}
\label{Analysis}

The experimental comparison gives credits to the model presented, and shows that the depth structure of the phenomenon is well reproduced. Starting from this point, the effect of the grains parameters (restitution and friction coefficients) and the fluid velocity fluctuations model are analyzed over a wide range of Shields numbers, in terms of dimensionless sediment transport rate versus Shields number, completed by solid depth profiles. The analysis aims at characterizing the influence of these parameters, in order to both study the influence of the different terms on the phenomenon, and the robustness of the experimental comparison. To extend the generality, a 3D bi-periodic (streamwise, spanwise) situation is considered. The random fixed bottom is generated from a gravity deposition, fixing all the particles contained in a slice of height $d$ at a given elevation in the granular bed. The size of the periodic cell has been chosen from a convergence analysis similar to the one undertaken for the 2D case (see appendix \ref{appendixAverage}) and a cell size of $l_x = l_y = 30d$ has been chosen to ensure statistical convergence and numerical stability. For each run, the DEM results are averaged over 100 seconds. \\

\subsection{Macroscopic considerations}
\begin{figure}
  \includegraphics[width=\textwidth]{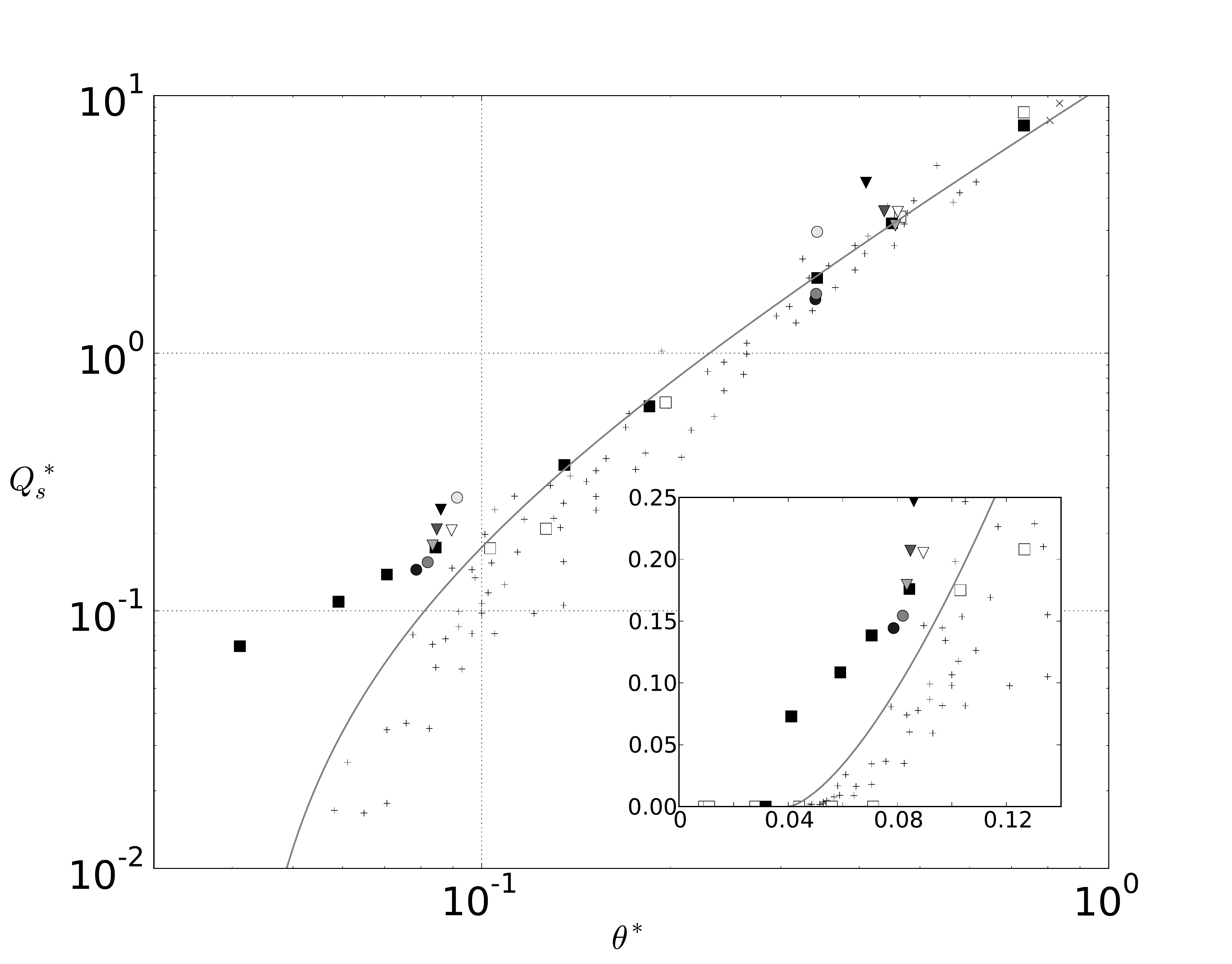}\\
\caption{\label{transportCurve}Dimensionless sediment transport rate $Q_s^*$ as a function of Shields number $\theta$ for different configuration. Classical runs ($\blacksquare$) with $e_n = 0.5$ and $\mu = 0.4$ are shown together with the exact same runs without turbulent fluctuation model ($\square$) for different Shields number. Triangle symbols represent the change in restitution coefficient with $e_n = 0.01$ ($\triangledown$),  $e_n = 0.25$ ($\color{grey1}{\blacktriangledown}$),  $e_n = 0.75$ ($\color{grey3}{\blacktriangledown}$), $e_n = 1$ ($\blacktriangledown$). Variations of the particle friction coefficient are represented by bullet points: $\mu=0.2$ ($\circ$),$\mu=0.6$ ($\color{grey3}{\bullet}$),$\mu=0.8$ ($\bullet$). The experimental data of Meyer-Peter and M\"uller \cite{MPM1948} (+) and Wilson \cite{Wilson1966} (x) synthesized in Yalin \cite{Yalin1977}, show the experimental trend in power law 3/2, with the dispersion of the data. The gray line corresponds to $Q_s^* = 11.8 (\theta^* - \theta_c^*)^{3/2}$ as found asymptotically by Wilson \cite{Wilson1987}. The inset in linear scale shows the behavior near the threshold of motion.}
\end{figure}

The dimensionless sediment transport rate as a function of the Shields number is presented in figure \ref{transportCurve}. The model results are compared  with experimental data from Meyer-Peter and M\"uller \cite{MPM1948} (+), and Wilson \cite{Wilson1966} (x). Simulation parameters for the reference configuration, represented as black squares ($\blacksquare$), are the same as the one used for the experimental comparison (see table \ref{tableInput}). The 3/2 power law is recovered by the numerical simulations and the results show a good agreement with experimental data for Shields number $\theta>0.1$. Near the threshold of motion, the model results differ from experimental measurements. The linear inset shows that the transition around the critical Shields number, characteristic of the onset of motion, is sharper in the numerical simulation results than in the experimental measurements. Also, the critical Shields number is slightly lower: around 0.04 in the model, against 0.047 for  Meyer-Peter and M\"uller \cite{MPM1948} data. The underestimation is consistent  with the use of spheres in the numerical simulations, which present smaller imbrication, and consequently smaller resistance to entrainment than the natural sediment used in the experiments. It is also worth noting that the scatter of the experimental data is usually very important close to the threshold of motion due to different definitions of the onset of motion and difficulties in shear stress measurements \cite{Buffington1997}. In particular, the present choice, based on the maximum turbulent shear stress, is less arbitrary than classical momentum balance estimates based on the water depth measurement, but most probably underestimates the Shields number. Considering the whole range of Shields number investigated, the results are in good agreement with literature data, and this shows that the numerical model is able to reproduce almost quantitatively the sediment transport rate.\\

The results of the model without the fluid velocity fluctuations are shown in figure \ref{transportCurve} as empty squares ($\square$). At high Shields number negligible differences are observed, while the influence is important close to the threshold of motion. It is consistent with the present conditions, where the suspension number is relatively high ($S^*=w_s/u_* \in [1.7 ; 10 ]$) and the fluid velocity fluctuations are expected to mostly play a role close to the threshold of motion. Focusing on the linear plot, it is observed that the critical Shields number is changed from around 0.04 to around 0.09 in the case without fluid velocity fluctuations. The former is in the range of  observed values under turbulent flow conditions \cite{Buffington1997}, while the latter is close to the value observed under laminar flow conditions \cite{Ouriemi2007}. The influence of the fluctuations on the critical Shields number can be associated with turbulent coherent structures (e.g. \cite{Papanicolaou2002,Dwivedi2012}). However, the present simple fluid velocity fluctuation model does not account for the space-time correlations induced by turbulent boundary layer coherent structures. This partly explains that the fluctuations model does not allow to describe well the evolution of the sediment transport rate with Shields number close to the threshold of motion (figure \ref{transportCurve}). Nevertheless it permits to successfully reproduce the onset of sediment transport motion in the turbulent regime, resulting in a good comparison with experimental depth profiles (section \ref{compResults}). \\

In the rigid grains limit, the granular interactions are characterized by the restitution and the friction coefficients. The restitution coefficient is representative of the energy loss during collisions, and has been shown experimentally to decrease with the impact velocity following a power law exponent lower than 1/4 \cite{Gondret2002}. For a limited range of impact velocity and in a first approximation it can be considered as constant. As the fluid flow model does not allow to resolve the fluid at the particle scale, the local lubrication effect is included in the effective restitution coefficient $e_n$. Following Gondret {\it et al.} \cite{Gondret2002}, the effective restitution coefficient depends on the local Stokes number comparing the grain inertia to fluid viscous forces: $St = \rho^p v d/(9\eta^f)$ where $\eta^f$ is the fluid dynamic viscosity, and $v$ is the impact velocity. In the region where the collisions are dominant, the Stokes number is of order $10^2-10^3$, corresponding to effective restitution coefficient in the range $e_n \in [0.6e_n^d-0.9e_n^d]$, respectively, where $e_n^d$ is the restitution coefficient for dry grains \cite{Gondret2002}. In our model a constant restitution coefficient is adopted, taking into account the lubrication effect globally. It is therefore a characteristic of the material and of the lubrication effect.\\

The influence of the restitution coefficient is shown in figure \ref{transportCurve} for two different Shields numbers, by keeping the free-surface position and number of particle layers constant. The restitution coefficient has been varied in the range $e_n \in [0.01,1]$. It corresponds to a realistic range $e_n \in [0.25,0.75]$, complemented by two extreme cases: no rebounds ($e_n = 0.01$) and no dissipation at contact ($e_n = 1$). Focusing on the realistic range at high Shields number ($\theta \sim 0.45$), the effect on the sediment transport rate is negligible. A slight trend is observed, the sediment transport rate and the Shields number being respectively increasing and decreasing function of the restitution coefficient. The extreme case without dissipation at contact ($e_n=1$) follows the same trend but exhibits a more important transport rate increase. Quite surprisingly, the case $e_n = 0.01$ shows an increase in transport rate with respect to case $e_n = 0.25$. For the lower Shields number value ($\theta \sim 0.1$), while the dependency in restitution coefficient is limited in the realistic range, there is no associated clear trend.\\The non-monotonous dependencies observed show non-trivial coupling between the granular phase characteristics and the sediment transport rate. The global weak dependency on the restitution coefficient is consistent with results obtained by Drake \& Calantoni \cite{Drake2001} under oscillatory flow conditions, and show that there is no need to include a lubrication model in the present condition. However, the relatively low importance of the restitution coefficient at such a high Shields number value is surprising. It is usually thought that collisional interactions are the dominant mechanism of momentum diffusion for such inertial particles \cite{Jenkins1998,Armanini2005}.\\

The effect of the particle friction coefficient is also shown in figure \ref{transportCurve}, represented by circles: $\mu = 0.2$ ($\color{grey1}{\bullet}$), $\mu = 0.6$ ($\color{grey3}{\bullet}$), $\mu = 0.8$ ($\bullet$). Unlike for the restitution coefficient, the trend observed is monotonous for all values, and similar at low ($\theta \sim 0.09$) and high Shields number ($\theta \sim 0.35$). The sediment transport rate and the Shields number decrease with increasing friction coefficient. The effect appears to be non-linear as the observed influence for a variation from $\mu = 0.2$ to $0.4$ is much greater than the one observed for a variation from $\mu = 0.4$ and $0.8$. This type of dependency is characteristic of dry dense granular flows \cite{DaCruz2005}. \\

As a partial conclusion, the present analysis shows that (i) the 3/2 power law for the sediment transport versus Shields number relationship is well captured by the proposed model; (ii) the fluid velocity fluctuations model is essential to capture a realistic value for the critical Shields number under turbulent flow conditions (iii) the influence of the granular interaction parameters is low, when taken in a realistic range. These results underline the robustness of the model and strengthen the experimental validation. Extreme value of particle friction and restitution coefficient affects the results, and show complex behaviors. In order to understand better the mechanisms at work, the sensitivity to granular interaction parameters will be further discussed by analyzing the results in terms of depth profiles.

\subsection{Depth profiles analysis}
\begin{figure}
  \includegraphics[width=\textwidth]{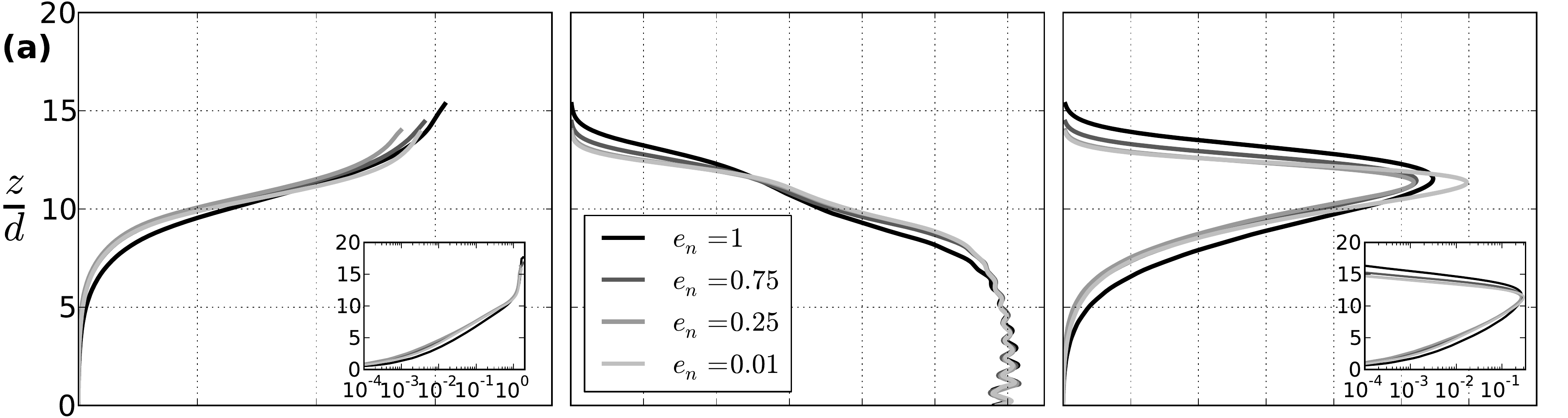}\\
  \includegraphics[width=\textwidth]{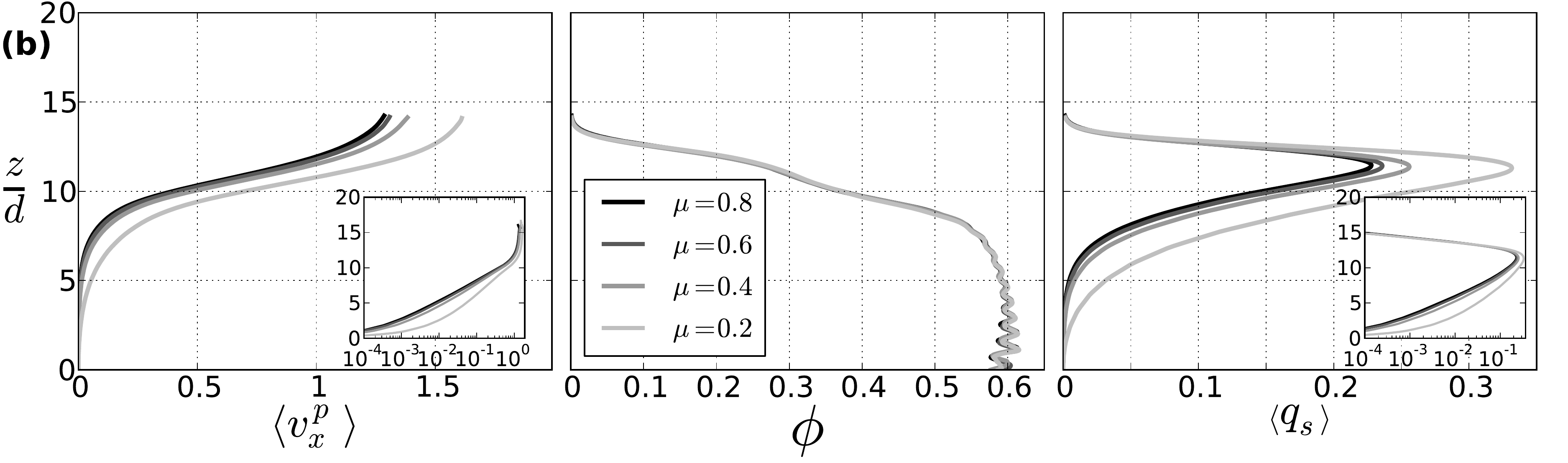}\\
\caption{\label{3DProfiles}Effect of the restitution (a) and friction coefficient (b) on the average solid depth profiles for a Shields number $\theta \sim 0.45$. The solid velocity $\left<v_x^p\right>$ and sediment transport rate density $\left<q_s\right>$ are given in $m/s$, while the solid volume fraction $\phi$ is dimensionless. To give a global picture of the trend, the color of the lines are representative of the friction and restitution coefficient values. The position of the free surface in both case is situated at $20d$. The curve are plotted only for values of solid volume fraction larger than $10^{-3}$.}
\end{figure}

Figure \ref{3DProfiles}a shows the solid depth profiles for a Shields number value $\theta \sim 0.45$, and for the different restitution coefficient values. Such a high Shields number value enhances the effect of restitution coefficient. From the profile, a clear trend appears, the sediment transport rate density profile is broader with increasing restitution coefficient. Excepting case $e_n = 0.01$, this is associated with an overall increase of particle velocity throughout the depth, the velocity profile being shifted with almost the same shape. The solid volume fraction profile shows an increase of the mobile layer thickness when the restitution coefficient is increased: the solid volume fraction is lowered close to the quasi-static bed while it is increased in the upper part of the flow. This can be explained considering predictions of the kinetic theory of granular flows, where particle phase normal stress is an increasing function of the restitution coefficient \cite{Jenkins1998}. The increase in particle normal stress is logically associated with a decompaction of the bed, which is submitted only to gravity. The case $e_n = 0.01$ is peculiar, the global trend is observed in the lower part of the flow while a higher particle velocity is predicted in the upper part. This reflects the coupling with the fluid phase, and features the complex mechanisms at work.\\
Figure \ref{3DProfiles}b shows the influence at high Shields number $\theta \sim 0.45$ of the friction coefficient over the range  $\mu \in  [0.2 ; 0.8]$ with $0.2$ steps. Interestingly, the solid volume fraction profile is not affected by the variation in friction coefficient. On the contrary, the particle velocity and thus the sediment transport rate density profiles are increasing when the friction coefficient is decreased. The increase of the velocity throughout the depth is mainly affecting the lower part of the sediment transport rate density profile, where the solid volume fraction is maximum. It corresponds to the denser part of the granular flow, for which the frictional interactions are dominant. \\

This analysis shows that, while affecting weakly the macroscopical results, the friction and restitution coefficient impact the depth structure of the granular flow differently. In addition, the non-monotonous behavior observed suggests the presence of non-trivial coupling between the solid and the fluid phases. \\

\section{Conclusions}
\label{conclusion}

The model presented is a step toward a description of the granular processes of steady bedload transport. By adapting the closures to this particular case, it has been shown that the model is able to reproduce the classical macroscopic validation in term of sediment transport rate and Shields number. In addition, an original detailed validation with existing bedload transport experiments has been performed, comparing simultaneous measurements of average solid velocity and volume fraction. The good agreement with experiments together with the rather low sensitivity of the results to the granular parameters show the relevancy and the robustness of the proposed model, which reproduces not only the evolution of the sediment transport rate as a function of the Shields number, but also the depth structure of the granular phase. \\
The influence of the different model contributions have been studied. In particular, the discrete random walk fluid velocity fluctuations model has been shown to be sufficient to reproduce the reduction of the critical Shields number due to turbulent fluctuations. A weak impact of the restitution and friction coefficients variations has been observed on the macroscopic sediment transport rate versus Shields number curve. The analysis of the depth profiles variations shows however that the granular parameters influence the depth structure of the granular flows and induce non-trivial coupling with the fluid phase. \\

The rigorous development of the model and the experimental validations demonstrate the potential of this modeling approach to deal with granular processes in bedload transport. Future work will take advantage of the description of the depth structure to analyze the effective granular rheology under bedload transport conditions.

\appendix
\label{appendix}

\section{Averaging expression and convergence analysis}
\label{appendixAverage}

The averaging of the solid phase takes a central part in the coupling between the Lagrangian solid phase and the Eulerian fluid phase. The important wall-normal gradient requires the length scale of the weighting function in this direction to be lower than particle diameter ($l_z \simeq d/30$ for the lowest Shields number) in order to define a rigorous averaging. We postulated that the complementary length scales $l_x$ and $l_y$ can statistically compensate the limited $l_z$. \\
In the present model, the average fluid description is 1D so that it depends only on the wall-normal component, $z$. The solid averaging can therefore be performed on the full width and length cell. With the cuboidal formulation of the weighting function defined previously (eq. \ref{weightingFunction}), the solid averaging of a scalar particle quantity $\gamma$ at a wall-normal position $z$ can be rewritten as:  
\begin{equation}
\left<\gamma\right>^s(z) = \frac{1}{\phi(z)} \sum_{ \{p | z^p \in [z- l_z/2 ; z+ l_z/2]\} } \tilde{V}^p \gamma^p
\label{averagingFormula}
\end{equation}
Where $l_z$ is the defined wall-normal weighting function length scale, and $\tilde{V}^p$ is the fraction of the particle volume contained in the slice of height $l_z$ at elevation $z$. We recover here the averaging formulation of Hill {\it et al} \cite{Hill2003}, which is convenient to compute since the volume of a slice of spheres can be evaluated analytically. The averaging box height $l_z$ is imposed by the vertical grid size of the fluid problem and no overlapping between the different slices is allowed.\\

For each fluid resolution, so at each given time step, the statistical representativity of the averages is a requirement for a consistent definition of the averaging process (section \ref{averageProc}). The spatial convergence of the averages with increasing complementary length scales $l_x$ and $l_y$ includes the effect of the bottom boundary conditions and particles arrangement, in addition to the statistical representativity. The results are required to be independent from the three effects, and consequently the spatial convergence of the results with respect to  $l_x$ and $l_y$ is analyzed in the present appendix.\\
There are two different convergence scale in the problem. The first one is associated to the spatial convergence at each given time step, and the second one is associated to the temporal convergence of a simulation with a given cell size. In the present analysis, as the paper focus on steady equilibrium results, we consider time-averaged results which are converged in time. The convergence analysis is conducted with respect to a large reference cell size for which we consider that the results at each fluid resolution are spatially converged. Indeed, a convergence with respect to this case ensures that the error made in the spatial averaging along the simulation are compensating each other. The analysis focuses on the transport rate depth profile. For both 2D and 3D cases, $l_z$ is taken at its minimal value in the problem $l_z = d/30$.\\

\paragraph*{quasi-2D convergence analysis}
We present here the results of the time-averaged spatial convergence analysis for the quasi-2D case Sim20 detailed in section \ref{appModel}. In this configuration $l_y$ is fixed at the channel width, and the problem is considered only as a function of the streamwise length $l_x$ which determines the size of the averaging cells. The convergence analysis is made with respect to the reference state chosen as $l_x = 10000d$, corresponding to a periodic length cell of $60m$ for particles of $6mm$ and about $80000$ particles in the simulation. We performed different simulations with a periodic streamwise length cell $l_x$ of respectively $50$, $100$, $200$, $300$, $500$, $1000$, $2500$ and $5000~d$.\\ 

To quantitatively analyze the transport rate profile differences, an indicator representative of the deviation with respect to a reference case is defined. It is given as the root mean square (RMS) of the difference between the considered transport rate profile and the reference one:
\begin{equation}
\displaystyle \frac{{{\Delta Q}_{rms}}^i}{\left<Q^{ref}\right>} = \frac{\sqrt{\frac{1}{N}\sum_{z = 0}^{N}{(\left<Q\right>_z^i - \left<Q\right>_z^{ref})^2}}}{\frac{1}{N}\sum_{z = 0}^{N}{\left<Q\right>_z^{ref}}},
\label{RMSerror}
\end{equation}
where the RMS ${{\Delta Q}_{rms}}^i$ is normalized by the average transport rate of the reference configuration  $\left<Q\right>^{ref}$,  $N$ is the number of averaging cell in the depth, $\left<Q\right>_z^i$ and $\left<Q\right>_z^{ref}$ are the values of transport rate in the cell z for the considered case and the reference case respectively. This variable effectively measures how close the results considered is from the reference case. \\

\begin{figure}
  \includegraphics[width=\textwidth]{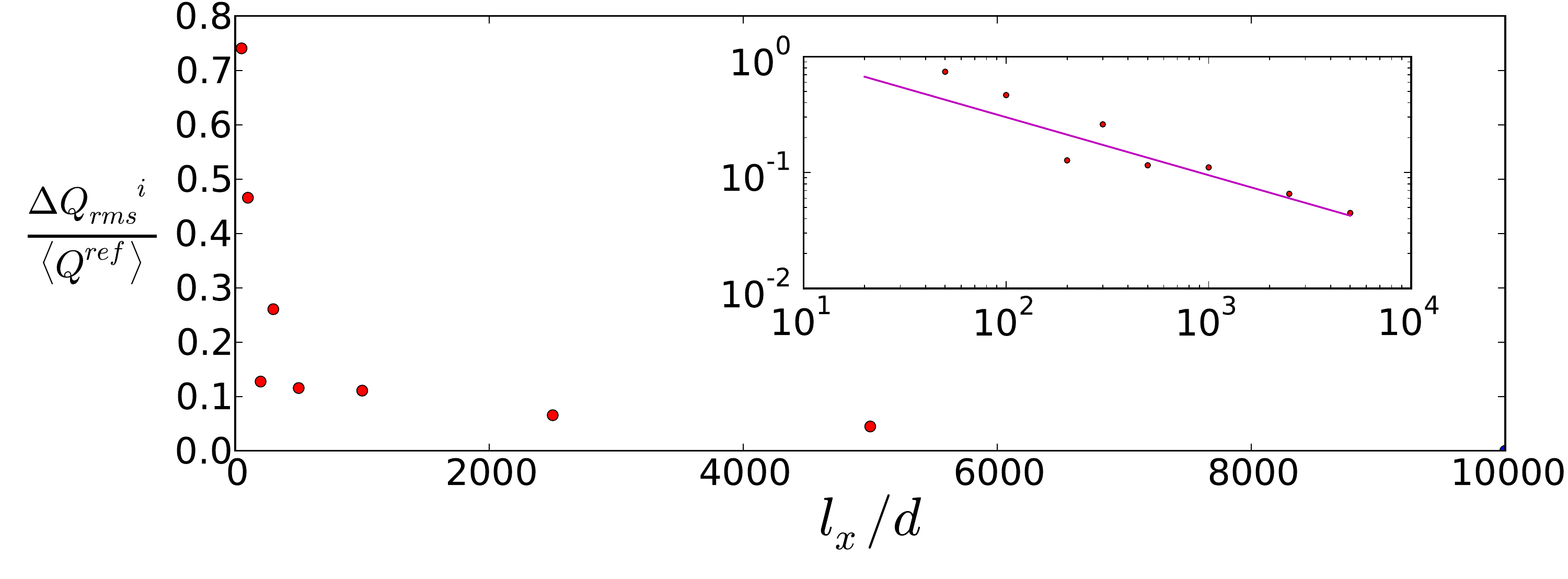}
\caption{\label{convergenceFig}Convergence of the average sediment transport rate profile as a function of the periodic cell size considered for the quasi-2D case. The vertical axis represents the deviation with respect to the reference configuration ($l_x = 10000d$) as defined in equation (\ref{RMSerror}). The inset shows that the convergence is slightly superior to $l_x^{-0.5}$ (\textcolor{magenta}{--}).}
\end{figure}
Figure \ref{convergenceFig} shows the normalized deviation with respect to the reference configuration defined by equation (\ref{RMSerror}) as a function of the streamwise periodic cell length of the simulation considered. The time-averaged results show a convergence as a function of cell length $l_x$ of the order of $l_x^{-\frac{1}{2}}$. The size of the periodic cell used for the simulations presented in the experimental comparison of the paper, was chosen as $l_x = 1000~d$, as it gives the best trade off between computational time and deviation observed.\\

\paragraph*{3D convergence analysis}
\begin{figure}
  \includegraphics[width=\textwidth]{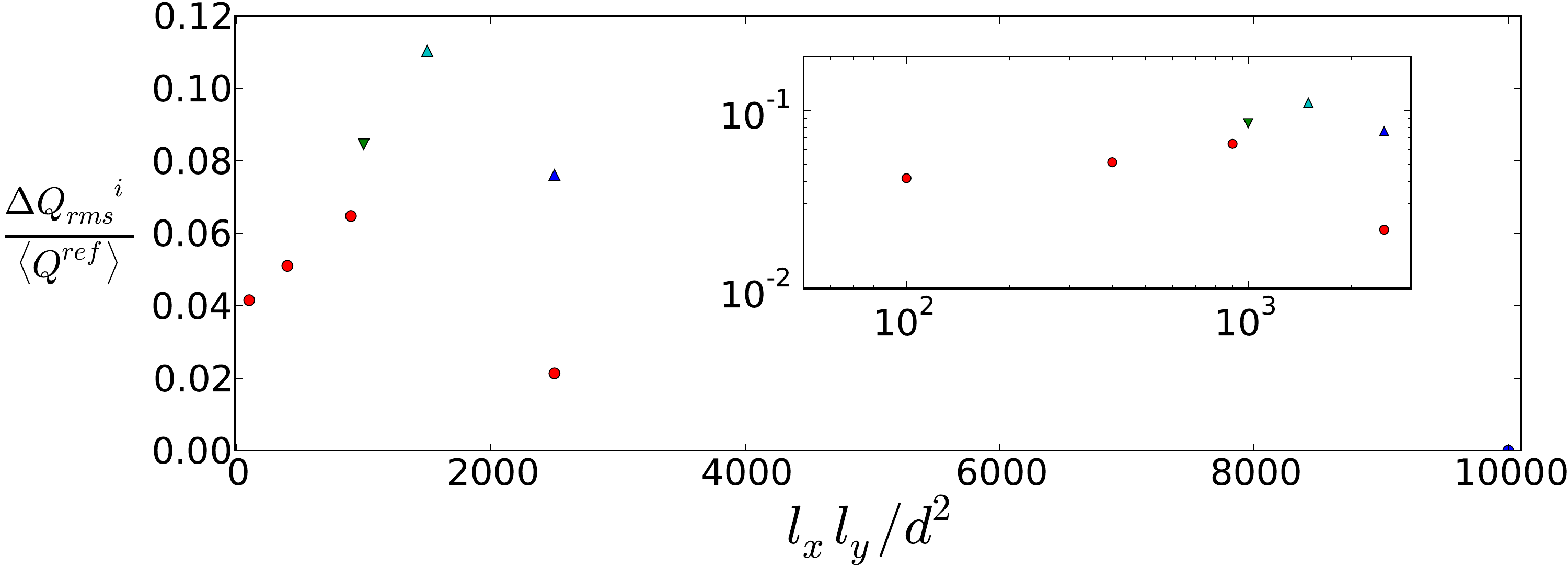}
\caption{\label{convergenceFigBIS}Convergence of the sediment transport rate profile as a function of the periodic cell size considered for the 3D case. The vertical axis represents the deviation with respect to the reference configuration ($l_x = l_y = 100d$) as defined in equation (\ref{RMSerror}). Cases with $l_x = l_y$ are represented with $\color{red}{\bullet}$, while the triangles denotes cases with $l_x \neq l_y$:  $(l_x,l_y) = (500~d,5~d)$ ($\color{blue}{\blacktriangle}$), $(300~d,5~d)$ ($\color{cyan}{\blacktriangle}$), and $(10~d,100~d)$ ($\color{green}{\blacktriangledown}$). The logarithmic scale inset shows that the results are already converged.}
\end{figure}
A similar analysis has been undertaken for the three dimensional bi-periodic configuration at a Shields number around $0.1$. The reference case has been chosen as $l_x = l_y = 100d$, i.e. the relative cell size $V = l_x~l_y~l_z$ being the same as the quasi-2D case. Different simulations have been performed with a cell of $l_x = l_y = 10~d$, $20~d$, $30~d$,$50~d$, and three cases with $l_x \neq l_y$: $(l_x,l_y) = (500~d,5~d)$, $(300~d,5~d)$, and $(10~d,100~d)$ . It has not been possible to consider smaller cell sizes, as the coupled model becomes unstable.\\
The main results are summarized in figure \ref{convergenceFigBIS}, expressing the RMS deviation with respect to the reference case as a function of the product $l_x~l_y/d^2$. The latter reflects the statistical representativity as it directly determines the size of the averaging cell $V = l_x~l_y~l_z$ ($l_z = dz$ fixed). The figure shows that cases with $l_x = l_y$ give better convergence than the ones with $l_x \neq l_y$.  There does not seem to be a convergence with increasing cell area. To our opinion, this reflects the fact that the results are already converged. Interestingly, the relative periodic cell size for convergence is less important in the 3D configuration, as a cell size of  $l_x = l_y = 10d$ (to compare with $l_x = 1000d$ and $l_y = 1d$) is almost already converged with respect to $l_x = l_y = 100d$. This can be explained by the better randomness of the 3D packing, and suggests that the statistical representativity was not the limiting parameter in the quasi-2D convergence analysis. For 3D cases, a higher Shields number increases the numerical instability of the coupling, so that it has been necessary to consider cell sizes up to $l_x = l_y = 50d$ for the highest Shields numbers simulations presented in the paper.

\section{fluid resolution period}
\label{appendixFluid}
\begin{table}
\caption{\label{tableFluidSensitivity}Results of the sensitivity analysis to the fluid resolution period $\tau_f$, for the case Sim20. The measured Shields number $\theta^*$, transport rate $\dot{n}$, and RMS deviation with respect to the case Ref. $\frac{{{\Delta Q}_{rms}}^i}{\left<Q^{ref}\right>}$, are given for each case.}
\begin{ruledtabular}
\begin{tabular}{ccccc}
Case & $\tau_f ~(s)$ & $\theta^*$ &  $\dot{n}~ (b/s)$ & $\frac{{{\Delta Q}_{rms}}^i}{\left<Q^{ref}\right>}$\\
\hline\noalign{\smallskip}
Ref. &   		$10^{-2}$ & 0.067 &  16.83 &  0.0 \\
$\tau_f= 10^{-3}$ & 	$10^{-3}$ & 0.067 &  16.74 &  0.06\\ 
$\tau_f = 10^{-1}$ & 	$10^{-1}$ & 0.067 &  16.91 &  0.04 \\
$\tau_f = 1$ & 		$1$	  & 0.066 &  17.26 &  0.13\\
\end{tabular}
\end{ruledtabular}
\end{table}

The DEM time step needs to be particularly low and the evolution of the granular medium over this time is limited. Consequently the fluid resolution period $\tau_D$ does not need necessarily to be equal to the solid time step. The stability of the coupling however depends on this period of resolution and it is important to study this parameter in order to have a meaningful model. The fluid resolution period should be defined to be smaller than the characteristic time of evolution of the granular medium. The fluid resolution is 1D and the equation is influenced only by the streamwise particle velocity and the wall-normal particle position (through respectively $\left<f^p_{D,x}\right>$ and $\phi$). For a single particle the evolution of these properties depends on the collisions and the entrainment by the fluid. As seen previously (section \ref{Analysis}), collisions do not significantly influence the phenomenon so that we consider only the characteristic time of entrainment. As explained in section \ref{coupling}, the characteristic time of relaxation to accelerate a particle to the fluid velocity is given by $\tau_D = \beta^{-1}$, with $\beta$ expressed from equation (\ref{beta}). In the present case it is of order $\tau_D \sim 10^{-1}s$. However, the characteristic time associated with each independent particle is not in general representative of the evolution of a complex many-body problem. \\

We therefore performed a sensitivity analysis on the period of fluid resolution. The results are shown in table \ref{tableFluidSensitivity} in term of RMS deviation (as defined by equation (\ref{RMSerror})) with respect to the reference configuration Sim20 for which $\tau_f = 10^{-2}s$. It includes $\tau_f = 10^{-3}s$, $10^{-1}s$, and $1s$. The results exhibit no dependence on the fluid period resolution in the range $10^{-3}s$ to $1s$. The values of the RMS deviation with respect to the reference configuration $\tau_f = 10^{-2}s$ is for all cases below $0.1$, i.e. below the reproducibility deviation value. The fluid resolution period therefore does not have an influence on the averaged equilibrium results within the range considered. This means that the solid average quantities transmitted to the fluid do not vary importantly during the simulation after reaching equilibrium. These results confirm that the simulations considered are at transport equilibrium. The fluid resolution period could however be important for unsteady conditions.

\begin{acknowledgments}
We are grateful to Michael Church for general remarks and English corrections. This research was supported by Irstea (formerly Cemagref) and the French Institut National des Sciences de l'Univers program EC2CO-BIOHEFECT, and EC2CO-LEFE « MODSED ».
\end{acknowledgments}

\bibliography{MaurinetalPoF_bibtexFile.bib}

\begin{thebibliography}{65}%
\makeatletter
\providecommand \@ifxundefined [1]{%
 \@ifx{#1\undefined}
}%
\providecommand \@ifnum [1]{%
 \ifnum #1\expandafter \@firstoftwo
 \else \expandafter \@secondoftwo
 \fi
}%
\providecommand \@ifx [1]{%
 \ifx #1\expandafter \@firstoftwo
 \else \expandafter \@secondoftwo
 \fi
}%
\providecommand \natexlab [1]{#1}%
\providecommand \enquote  [1]{``#1''}%
\providecommand \bibnamefont  [1]{#1}%
\providecommand \bibfnamefont [1]{#1}%
\providecommand \citenamefont [1]{#1}%
\providecommand \href@noop [0]{\@secondoftwo}%
\providecommand \href [0]{\begingroup \@sanitize@url \@href}%
\providecommand \@href[1]{\@@startlink{#1}\@@href}%
\providecommand \@@href[1]{\endgroup#1\@@endlink}%
\providecommand \@sanitize@url [0]{\catcode `\\12\catcode `\$12\catcode
  `\&12\catcode `\#12\catcode `\^12\catcode `\_12\catcode `\%12\relax}%
\providecommand \@@startlink[1]{}%
\providecommand \@@endlink[0]{}%
\providecommand \url  [0]{\begingroup\@sanitize@url \@url }%
\providecommand \@url [1]{\endgroup\@href {#1}{\urlprefix }}%
\providecommand \urlprefix  [0]{URL }%
\providecommand \Eprint [0]{\href }%
\providecommand \doibase [0]{http://dx.doi.org/}%
\providecommand \selectlanguage [0]{\@gobble}%
\providecommand \bibinfo  [0]{\@secondoftwo}%
\providecommand \bibfield  [0]{\@secondoftwo}%
\providecommand \translation [1]{[#1]}%
\providecommand \BibitemOpen [0]{}%
\providecommand \bibitemStop [0]{}%
\providecommand \bibitemNoStop [0]{.\EOS\space}%
\providecommand \EOS [0]{\spacefactor3000\relax}%
\providecommand \BibitemShut  [1]{\csname bibitem#1\endcsname}%
\let\auto@bib@innerbib\@empty
\bibitem [{\citenamefont {DuBoys}(1879)}]{DuBoys1879}%
  \BibitemOpen
  \bibfield  {author} {\bibinfo {author} {\bibfnamefont {M.~P.}\ \bibnamefont
  {DuBoys}},\ }\bibfield  {title} {\enquote {\bibinfo {title} {Le rh\^one et
  les rivi\`eres \`a lit affouillable},}\ }\href@noop {} {\bibfield  {journal}
  {\bibinfo  {journal} {Annales des Ponts et Chauss\'ees}\ }\textbf {\bibinfo
  {volume} {18}},\ \bibinfo {pages} {141--195} (\bibinfo {year}
  {1879})}\BibitemShut {NoStop}%
\bibitem [{\citenamefont {Meyer-Peter}\ and\ \citenamefont
  {M\"uller}(1948)}]{MPM1948}%
  \BibitemOpen
  \bibfield  {author} {\bibinfo {author} {\bibfnamefont {E.}~\bibnamefont
  {Meyer-Peter}}\ and\ \bibinfo {author} {\bibfnamefont {R.}~\bibnamefont
  {M\"uller}},\ }\bibfield  {title} {\enquote {\bibinfo {title} {Formulas for
  bed-load transport},}\ }in\ \href@noop {} {\emph {\bibinfo {booktitle} {Proc.
  2nd Meeting}}}\ (\bibinfo  {publisher} {IAHR},\ \bibinfo {year} {1948})\ pp.\
  \bibinfo {pages} {39--64}\BibitemShut {NoStop}%
\bibitem [{\citenamefont {Bathurst}(2007)}]{Bathurst2007}%
  \BibitemOpen
  \bibfield  {author} {\bibinfo {author} {\bibfnamefont {J.}~\bibnamefont
  {Bathurst}},\ }\bibfield  {title} {\enquote {\bibinfo {title} {Effect of
  coarse surface layer on bed-load transport},}\ }\href {\doibase
  10.1061/(ASCE)0733-9429(2007)133:11(1192)} {\bibfield  {journal} {\bibinfo
  {journal} {Journal of Hydraulic Engineering}\ }\textbf {\bibinfo {volume}
  {133}},\ \bibinfo {pages} {1192--1205} (\bibinfo {year} {2007})}\BibitemShut
  {NoStop}%
\bibitem [{\citenamefont {Recking}(2010)}]{Recking2010}%
  \BibitemOpen
  \bibfield  {author} {\bibinfo {author} {\bibfnamefont {A.}~\bibnamefont
  {Recking}},\ }\bibfield  {title} {\enquote {\bibinfo {title} {A comparison
  between flume and field bed load transport data and consequences for
  surface-based bed load transport prediction},}\ }\href {\doibase
  10.1029/2009WR008007} {\bibfield  {journal} {\bibinfo  {journal} {Water
  Resources Research}\ }\textbf {\bibinfo {volume} {46}},\ \bibinfo {pages}
  {W03518} (\bibinfo {year} {2010})}\BibitemShut {NoStop}%
\bibitem [{\citenamefont {Jackson}(2000)}]{Jackson2000}%
  \BibitemOpen
  \bibfield  {author} {\bibinfo {author} {\bibfnamefont {R.}~\bibnamefont
  {Jackson}},\ }\href@noop {} {\emph {\bibinfo {title} {The dynamics of
  fluidized particles}}}\ (\bibinfo  {publisher} {Cambridge University Press},\
  \bibinfo {year} {2000})\BibitemShut {NoStop}%
\bibitem [{\citenamefont {Drew}\ and\ \citenamefont
  {Passman}(1999)}]{Drew1999}%
  \BibitemOpen
  \bibfield  {author} {\bibinfo {author} {\bibfnamefont {D.~A.}\ \bibnamefont
  {Drew}}\ and\ \bibinfo {author} {\bibfnamefont {S.~L.}\ \bibnamefont
  {Passman}},\ }\href@noop {} {\emph {\bibinfo {title} {Theory of
  Multicomponent Fluids}}}\ (\bibinfo  {publisher} {Springer},\ \bibinfo {year}
  {1999})\BibitemShut {NoStop}%
\bibitem [{\citenamefont {Hanes}\ and\ \citenamefont
  {Bowen}(1985)}]{Hanes1985}%
  \BibitemOpen
  \bibfield  {author} {\bibinfo {author} {\bibfnamefont {D.~M.}\ \bibnamefont
  {Hanes}}\ and\ \bibinfo {author} {\bibfnamefont {A.~J.}\ \bibnamefont
  {Bowen}},\ }\bibfield  {title} {\enquote {\bibinfo {title} {A granular-fluid
  model for steady intense bed-load transport},}\ }\href {\doibase
  10.1029/JC090iC05p09149} {\bibfield  {journal} {\bibinfo  {journal} {Journal
  of Geophysical Research: Oceans}\ }\textbf {\bibinfo {volume} {90}},\
  \bibinfo {pages} {9149--9158} (\bibinfo {year} {1985})}\BibitemShut {NoStop}%
\bibitem [{\citenamefont {Revil-Baudard}\ and\ \citenamefont
  {Chauchat}(2013)}]{RevilBaudard2013}%
  \BibitemOpen
  \bibfield  {author} {\bibinfo {author} {\bibfnamefont {T.}~\bibnamefont
  {Revil-Baudard}}\ and\ \bibinfo {author} {\bibfnamefont {J.}~\bibnamefont
  {Chauchat}},\ }\bibfield  {title} {\enquote {\bibinfo {title} {A two-phase
  model for sheet flow regime based on dense granular flow rheology},}\ }\href
  {\doibase 10.1029/2012JC008306} {\bibfield  {journal} {\bibinfo  {journal}
  {Journal of Geophysical Research: Oceans}\ }\textbf {\bibinfo {volume}
  {118}},\ \bibinfo {pages} {619--634} (\bibinfo {year} {2013})}\BibitemShut
  {NoStop}%
\bibitem [{\citenamefont {Aussillous}\ \emph {et~al.}(2013)\citenamefont
  {Aussillous}, \citenamefont {Chauchat}, \citenamefont {Pailha}, \citenamefont
  {M\'edale},\ and\ \citenamefont {Guazzelli}}]{Aussillous2013}%
  \BibitemOpen
  \bibfield  {author} {\bibinfo {author} {\bibfnamefont {P.}~\bibnamefont
  {Aussillous}}, \bibinfo {author} {\bibfnamefont {J.}~\bibnamefont
  {Chauchat}}, \bibinfo {author} {\bibfnamefont {M.}~\bibnamefont {Pailha}},
  \bibinfo {author} {\bibfnamefont {M.}~\bibnamefont {M\'edale}}, \ and\
  \bibinfo {author} {\bibfnamefont {E.}~\bibnamefont {Guazzelli}},\ }\bibfield
  {title} {\enquote {\bibinfo {title} {Investigation of the mobile granular
  layer in bedload transport by laminar shearing flows},}\ }\href {\doibase
  10.1017/jfm.2013.546} {\bibfield  {journal} {\bibinfo  {journal} {Journal of
  Fluid Mechanics}\ }\textbf {\bibinfo {volume} {736}},\ \bibinfo {pages}
  {594--615} (\bibinfo {year} {2013})}\BibitemShut {NoStop}%
\bibitem [{\citenamefont {Jenkins}\ and\ \citenamefont
  {Hanes}(1998)}]{Jenkins1998}%
  \BibitemOpen
  \bibfield  {author} {\bibinfo {author} {\bibfnamefont {J.}~\bibnamefont
  {Jenkins}}\ and\ \bibinfo {author} {\bibfnamefont {D.}~\bibnamefont
  {Hanes}},\ }\bibfield  {title} {\enquote {\bibinfo {title} {Collisional sheet
  flows of sediment driven by a turbulent fluid.}}\ }\href {\doibase
  10.1017/S0022112098001840} {\bibfield  {journal} {\bibinfo  {journal}
  {Journal of Fluid Mechanics}\ }\textbf {\bibinfo {volume} {370}},\ \bibinfo
  {pages} {29--52} (\bibinfo {year} {1998})}\BibitemShut {NoStop}%
\bibitem [{\citenamefont {Hsu}, \citenamefont {Jenkins},\ and\ \citenamefont
  {Liu}(2004)}]{Hsu2004}%
  \BibitemOpen
  \bibfield  {author} {\bibinfo {author} {\bibfnamefont {T.}~\bibnamefont
  {Hsu}}, \bibinfo {author} {\bibfnamefont {J.}~\bibnamefont {Jenkins}}, \ and\
  \bibinfo {author} {\bibfnamefont {P.}~\bibnamefont {Liu}},\ }\bibfield
  {title} {\enquote {\bibinfo {title} {On two-phase sediment transport: sheet
  flow of massive particles},}\ }\href {\doibase 10.1098/rspa.2003.1273}
  {\bibfield  {journal} {\bibinfo  {journal} {Proc. of the Royal Society of
  London A}\ }\textbf {\bibinfo {volume} {460}},\ \bibinfo {pages} {2223--2250}
  (\bibinfo {year} {2004})}\BibitemShut {NoStop}%
\bibitem [{\citenamefont {Tsuji}, \citenamefont {Kawaguchi},\ and\
  \citenamefont {Tanaka}(1993)}]{Tsuji1993}%
  \BibitemOpen
  \bibfield  {author} {\bibinfo {author} {\bibfnamefont {Y.}~\bibnamefont
  {Tsuji}}, \bibinfo {author} {\bibfnamefont {T.}~\bibnamefont {Kawaguchi}}, \
  and\ \bibinfo {author} {\bibfnamefont {T.}~\bibnamefont {Tanaka}},\
  }\bibfield  {title} {\enquote {\bibinfo {title} {Discrete particle simulation
  of two-dimensional fluidized bed},}\ }\href {\doibase
  10.1016/0032-5910(93)85010-7} {\bibfield  {journal} {\bibinfo  {journal}
  {Powder Technology}\ }\textbf {\bibinfo {volume} {77}},\ \bibinfo {pages} {79
  -- 87} (\bibinfo {year} {1993})}\BibitemShut {NoStop}%
\bibitem [{\citenamefont {Zhu}\ \emph {et~al.}(2007)\citenamefont {Zhu},
  \citenamefont {Zhou}, \citenamefont {Yang},\ and\ \citenamefont
  {Yu}}]{Zhu2007}%
  \BibitemOpen
  \bibfield  {author} {\bibinfo {author} {\bibfnamefont {H.}~\bibnamefont
  {Zhu}}, \bibinfo {author} {\bibfnamefont {Z.}~\bibnamefont {Zhou}}, \bibinfo
  {author} {\bibfnamefont {R.}~\bibnamefont {Yang}}, \ and\ \bibinfo {author}
  {\bibfnamefont {A.}~\bibnamefont {Yu}},\ }\bibfield  {title} {\enquote
  {\bibinfo {title} {Discrete particle simulation of particulate systems:
  Theoretical developments},}\ }\href {\doibase 10.1016/j.ces.2006.12.089}
  {\bibfield  {journal} {\bibinfo  {journal} {Chemical Engineering Science}\
  }\textbf {\bibinfo {volume} {62}},\ \bibinfo {pages} {3378 -- 3396} (\bibinfo
  {year} {2007})}\BibitemShut {NoStop}%
\bibitem [{\citenamefont {Kidanemariam}\ \emph {et~al.}(2013)\citenamefont
  {Kidanemariam}, \citenamefont {Chan-Braun}, \citenamefont {Doychev},\ and\
  \citenamefont {Uhlmann}}]{Kidanemariam2013}%
  \BibitemOpen
  \bibfield  {author} {\bibinfo {author} {\bibfnamefont {A.~G.}\ \bibnamefont
  {Kidanemariam}}, \bibinfo {author} {\bibfnamefont {C.}~\bibnamefont
  {Chan-Braun}}, \bibinfo {author} {\bibfnamefont {T.}~\bibnamefont {Doychev}},
  \ and\ \bibinfo {author} {\bibfnamefont {M.}~\bibnamefont {Uhlmann}},\
  }\bibfield  {title} {\enquote {\bibinfo {title} {Direct numerical simulation
  of horizontal open channel flow with finite-size, heavy particles at low
  solid volume fraction},}\ }\href
  {http://stacks.iop.org/1367-2630/15/i=2/a=025031} {\bibfield  {journal}
  {\bibinfo  {journal} {New Journal of Physics}\ }\textbf {\bibinfo {volume}
  {15}},\ \bibinfo {pages} {025031} (\bibinfo {year} {2013})}\BibitemShut
  {NoStop}%
\bibitem [{\citenamefont {Jiang}\ and\ \citenamefont {Haff}(1993)}]{Jiang1993}%
  \BibitemOpen
  \bibfield  {author} {\bibinfo {author} {\bibfnamefont {Z.}~\bibnamefont
  {Jiang}}\ and\ \bibinfo {author} {\bibfnamefont {P.~K.}\ \bibnamefont
  {Haff}},\ }\bibfield  {title} {\enquote {\bibinfo {title} {Multiparticle
  simulation methods applied to the micromechanics of bed load transport},}\
  }\href {\doibase 10.1029/92WR02063} {\bibfield  {journal} {\bibinfo
  {journal} {Water Resources Research}\ }\textbf {\bibinfo {volume} {29}},\
  \bibinfo {pages} {399--412} (\bibinfo {year} {1993})}\BibitemShut {NoStop}%
\bibitem [{\citenamefont {Yeganeh}, \citenamefont {Gotoh},\ and\ \citenamefont
  {Sakai}(2000)}]{Yeganeh2000}%
  \BibitemOpen
  \bibfield  {author} {\bibinfo {author} {\bibfnamefont {A.}~\bibnamefont
  {Yeganeh}}, \bibinfo {author} {\bibfnamefont {H.}~\bibnamefont {Gotoh}}, \
  and\ \bibinfo {author} {\bibfnamefont {T.}~\bibnamefont {Sakai}},\ }\bibfield
   {title} {\enquote {\bibinfo {title} {Applicability of euler-lagrange
  coupling multiphase-flow model to bed-load transport under high bottom
  shear},}\ }\href {\doibase 10.1080/00221680009498320} {\bibfield  {journal}
  {\bibinfo  {journal} {Journal of Hydraulic Research}\ }\textbf {\bibinfo
  {volume} {38}},\ \bibinfo {pages} {389--398} (\bibinfo {year}
  {2000})}\BibitemShut {NoStop}%
\bibitem [{\citenamefont {Drake}\ and\ \citenamefont
  {Calantoni}(2001)}]{Drake2001}%
  \BibitemOpen
  \bibfield  {author} {\bibinfo {author} {\bibfnamefont {T.~G.}\ \bibnamefont
  {Drake}}\ and\ \bibinfo {author} {\bibfnamefont {J.}~\bibnamefont
  {Calantoni}},\ }\bibfield  {title} {\enquote {\bibinfo {title} {Discrete
  particle model for sheet flow sediment transport in the nearshore},}\ }\href
  {\doibase 10.1029/2000JC000611} {\bibfield  {journal} {\bibinfo  {journal}
  {J. Geophys. Res.}\ }\textbf {\bibinfo {volume} {106}},\ \bibinfo {pages}
  {19859--19868} (\bibinfo {year} {2001})}\BibitemShut {NoStop}%
\bibitem [{\citenamefont {Calantoni}, \citenamefont {Puleo},\ and\
  \citenamefont {Holland}(2006)}]{Calantoni2006}%
  \BibitemOpen
  \bibfield  {author} {\bibinfo {author} {\bibfnamefont {J.}~\bibnamefont
  {Calantoni}}, \bibinfo {author} {\bibfnamefont {J.~A.}\ \bibnamefont
  {Puleo}}, \ and\ \bibinfo {author} {\bibfnamefont {K.~T.}\ \bibnamefont
  {Holland}},\ }\bibfield  {title} {\enquote {\bibinfo {title} {Simulation of
  sediment motions using a discrete particle model in the inner surf and
  swash-zones},}\ }\href {\doibase 10.1016/j.csr.2005.11.013} {\bibfield
  {journal} {\bibinfo  {journal} {Continental Shelf Research}\ }\textbf
  {\bibinfo {volume} {26}},\ \bibinfo {pages} {610 -- 621} (\bibinfo {year}
  {2006})}\BibitemShut {NoStop}%
\bibitem [{\citenamefont {Yeganeh-Bakhtiary}\ \emph {et~al.}(2009)\citenamefont
  {Yeganeh-Bakhtiary}, \citenamefont {Shabani}, \citenamefont {Gotoh},\ and\
  \citenamefont {Wang}}]{Yeganeh-Bakhtiary2009}%
  \BibitemOpen
  \bibfield  {author} {\bibinfo {author} {\bibfnamefont {A.}~\bibnamefont
  {Yeganeh-Bakhtiary}}, \bibinfo {author} {\bibfnamefont {B.}~\bibnamefont
  {Shabani}}, \bibinfo {author} {\bibfnamefont {H.}~\bibnamefont {Gotoh}}, \
  and\ \bibinfo {author} {\bibfnamefont {S.~S.}\ \bibnamefont {Wang}},\
  }\bibfield  {title} {\enquote {\bibinfo {title} {A three-dimensional distinct
  element model for bed-load transport},}\ }\href {\doibase
  10.3826/jhr.2009.3168} {\bibfield  {journal} {\bibinfo  {journal} {Journal of
  Hydraulic Research}\ }\textbf {\bibinfo {volume} {47}},\ \bibinfo {pages}
  {203--212} (\bibinfo {year} {2009})}\BibitemShut {NoStop}%
\bibitem [{\citenamefont {Kidanemariam}\ and\ \citenamefont
  {Uhlmann}(2014)}]{Kidanemariam2014}%
  \BibitemOpen
  \bibfield  {author} {\bibinfo {author} {\bibfnamefont {A.~G.}\ \bibnamefont
  {Kidanemariam}}\ and\ \bibinfo {author} {\bibfnamefont {M.}~\bibnamefont
  {Uhlmann}},\ }\bibfield  {title} {\enquote {\bibinfo {title}
  {Interface-resolved direct numerical simulation of the erosion of a sediment
  bed sheared by laminar channel flow},}\ }\href {\doibase
  10.1016/j.ijmultiphaseflow.2014.08.008} {\bibfield  {journal} {\bibinfo
  {journal} {International Journal of Multiphase Flow}\ }\textbf {\bibinfo
  {volume} {67}},\ \bibinfo {pages} {174 -- 188} (\bibinfo {year}
  {2014})}\BibitemShut {NoStop}%
\bibitem [{\citenamefont {Duran}, \citenamefont {Andreotti},\ and\
  \citenamefont {Claudin}(2012)}]{Duran2012}%
  \BibitemOpen
  \bibfield  {author} {\bibinfo {author} {\bibfnamefont {O.}~\bibnamefont
  {Duran}}, \bibinfo {author} {\bibfnamefont {B.}~\bibnamefont {Andreotti}}, \
  and\ \bibinfo {author} {\bibfnamefont {P.}~\bibnamefont {Claudin}},\
  }\bibfield  {title} {\enquote {\bibinfo {title} {Numerical simulation of
  turbulent sediment transport, from bed load to saltation},}\ }\href {\doibase
  10.1063/1.4757662} {\bibfield  {journal} {\bibinfo  {journal} {Physics of
  Fluids}\ }\textbf {\bibinfo {volume} {24}},\ \bibinfo {eid} {103306}
  (\bibinfo {year} {2012})}\BibitemShut {NoStop}%
\bibitem [{\citenamefont {Ji}\ \emph {et~al.}(2013)\citenamefont {Ji},
  \citenamefont {Munjiza}, \citenamefont {Avital}, \citenamefont {Ma},\ and\
  \citenamefont {Williams}}]{Ji2013}%
  \BibitemOpen
  \bibfield  {author} {\bibinfo {author} {\bibfnamefont {C.}~\bibnamefont
  {Ji}}, \bibinfo {author} {\bibfnamefont {A.}~\bibnamefont {Munjiza}},
  \bibinfo {author} {\bibfnamefont {E.}~\bibnamefont {Avital}}, \bibinfo
  {author} {\bibfnamefont {J.}~\bibnamefont {Ma}}, \ and\ \bibinfo {author}
  {\bibfnamefont {J.~J.~R.}\ \bibnamefont {Williams}},\ }\bibfield  {title}
  {\enquote {\bibinfo {title} {Direct numerical simulation of sediment
  entrainment in turbulent channel flow},}\ }\href {\doibase 10.1063/1.4807075}
  {\bibfield  {journal} {\bibinfo  {journal} {Physics of Fluids}\ }\textbf
  {\bibinfo {volume} {25}},\ \bibinfo {eid} {056601} (\bibinfo {year}
  {2013})}\BibitemShut {NoStop}%
\bibitem [{\citenamefont {Fukuoka}, \citenamefont {Fukuda},\ and\ \citenamefont
  {Uchida}(2014)}]{Fukuoka2014}%
  \BibitemOpen
  \bibfield  {author} {\bibinfo {author} {\bibfnamefont {S.}~\bibnamefont
  {Fukuoka}}, \bibinfo {author} {\bibfnamefont {T.}~\bibnamefont {Fukuda}}, \
  and\ \bibinfo {author} {\bibfnamefont {T.}~\bibnamefont {Uchida}},\
  }\bibfield  {title} {\enquote {\bibinfo {title} {Effects of sizes and shapes
  of gravel particles on sediment transports and bed variations in a numerical
  movable-bed channel},}\ }\href {\doibase 10.1016/j.advwatres.2014.05.013}
  {\bibfield  {journal} {\bibinfo  {journal} {Advances in Water Resources}\
  }\textbf {\bibinfo {volume} {72}},\ \bibinfo {pages} {84 -- 96} (\bibinfo
  {year} {2014})}\BibitemShut {NoStop}%
\bibitem [{\citenamefont {Frey}\ and\ \citenamefont {Church}(2009)}]{Frey2009}%
  \BibitemOpen
  \bibfield  {author} {\bibinfo {author} {\bibfnamefont {P.}~\bibnamefont
  {Frey}}\ and\ \bibinfo {author} {\bibfnamefont {M.}~\bibnamefont {Church}},\
  }\bibfield  {title} {\enquote {\bibinfo {title} {How river beds move},}\
  }\href {\doibase 10.1126/science.1178516} {\bibfield  {journal} {\bibinfo
  {journal} {Science}\ }\textbf {\bibinfo {volume} {325}},\ \bibinfo {pages}
  {1509--1510} (\bibinfo {year} {2009})}\BibitemShut {NoStop}%
\bibitem [{\citenamefont {Frey}\ and\ \citenamefont {Church}(2010)}]{Frey2010}%
  \BibitemOpen
  \bibfield  {author} {\bibinfo {author} {\bibfnamefont {P.}~\bibnamefont
  {Frey}}\ and\ \bibinfo {author} {\bibfnamefont {M.}~\bibnamefont {Church}},\
  }\bibfield  {title} {\enquote {\bibinfo {title} {Bedload : a granular
  phenomenon},}\ }\href {\doibase 10.1002/esp.2103} {\bibfield  {journal}
  {\bibinfo  {journal} {Earth Surface Processes and Landform}\ }\textbf
  {\bibinfo {volume} {36}},\ \bibinfo {pages} {58--69} (\bibinfo {year}
  {2010})}\BibitemShut {NoStop}%
\bibitem [{\citenamefont {Kawaguchi}(2010)}]{Kawaguchi2010}%
  \BibitemOpen
  \bibfield  {author} {\bibinfo {author} {\bibfnamefont {T.}~\bibnamefont
  {Kawaguchi}},\ }\bibfield  {title} {\enquote {\bibinfo {title} {Mri
  measurement of granular flows and fluid-particle flows},}\ }\href {\doibase
  10.1016/j.apt.2010.03.014} {\bibfield  {journal} {\bibinfo  {journal}
  {Advanced Powder Technology}\ }\textbf {\bibinfo {volume} {21}},\ \bibinfo
  {pages} {235 -- 241} (\bibinfo {year} {2010})}\BibitemShut {NoStop}%
\bibitem [{\citenamefont {Frey}(2014)}]{Frey2014}%
  \BibitemOpen
  \bibfield  {author} {\bibinfo {author} {\bibfnamefont {P.}~\bibnamefont
  {Frey}},\ }\bibfield  {title} {\enquote {\bibinfo {title} {Particle velocity
  and concentration profiles in bedload experiments on a steep slope},}\ }\href
  {\doibase 10.1002/esp.3517} {\bibfield  {journal} {\bibinfo  {journal} {Earth
  Surface Processes and Landforms}\ }\textbf {\bibinfo {volume} {39}},\
  \bibinfo {pages} {646--655} (\bibinfo {year} {2014})}\BibitemShut {NoStop}%
\bibitem [{\citenamefont {Cundall}\ and\ \citenamefont
  {Strack}(1979)}]{Cundall1979}%
  \BibitemOpen
  \bibfield  {author} {\bibinfo {author} {\bibfnamefont {P.~A.}\ \bibnamefont
  {Cundall}}\ and\ \bibinfo {author} {\bibfnamefont {O.~D.~L.}\ \bibnamefont
  {Strack}},\ }\bibfield  {title} {\enquote {\bibinfo {title} {A discrete
  numerical model for granular assemblies},}\ }\href {\doibase
  10.1680/geot.1979.29.1.47} {\bibfield  {journal} {\bibinfo  {journal}
  {G\'eotechnique}\ ,\ \bibinfo {pages} {305--329(24)}} (\bibinfo {year}
  {1979})}\BibitemShut {NoStop}%
\bibitem [{\citenamefont {Li}\ and\ \citenamefont {Sawamoto}(1995)}]{Li1995}%
  \BibitemOpen
  \bibfield  {author} {\bibinfo {author} {\bibfnamefont {L.}~\bibnamefont
  {Li}}\ and\ \bibinfo {author} {\bibfnamefont {M.}~\bibnamefont {Sawamoto}},\
  }\bibfield  {title} {\enquote {\bibinfo {title} {Multi-phase model on
  sediment transport in sheet-flow regime under oscillatory flow},}\
  }\href@noop {} {\bibfield  {journal} {\bibinfo  {journal} {Coastal
  engineering Japan}\ }\textbf {\bibinfo {volume} {38}},\ \bibinfo {pages}
  {157--178} (\bibinfo {year} {1995})}\BibitemShut {NoStop}%
\bibitem [{\citenamefont {Prandtl}(1926)}]{Prandtl1926}%
  \BibitemOpen
  \bibfield  {author} {\bibinfo {author} {\bibfnamefont {L.}~\bibnamefont
  {Prandtl}},\ }\bibfield  {title} {\enquote {\bibinfo {title} {Bericht
  {\"u}ber neuere \textsc{T}urbulenzforschung},}\ }\href@noop {} {\bibfield
  {journal} {\bibinfo  {journal} {Hydraulische Probleme. Vortr{\"a}ge
  Hydrauliktagung G{\"o}ttingen}\ }\textbf {\bibinfo {volume} {5}},\ \bibinfo
  {pages} {1--13} (\bibinfo {year} {1926})}\BibitemShut {NoStop}%
\bibitem [{\citenamefont {Anderson}\ and\ \citenamefont
  {Jackson}(1967)}]{Anderson1967}%
  \BibitemOpen
  \bibfield  {author} {\bibinfo {author} {\bibfnamefont {T.~B.}\ \bibnamefont
  {Anderson}}\ and\ \bibinfo {author} {\bibfnamefont {R.}~\bibnamefont
  {Jackson}},\ }\bibfield  {title} {\enquote {\bibinfo {title} {Fluid
  mechanical description of fluidized beds. equations of motion},}\ }\href
  {\doibase 10.1021/i160024a007} {\bibfield  {journal} {\bibinfo  {journal}
  {Industrial \& Engineering Chemistry Fundamentals}\ }\textbf {\bibinfo
  {volume} {6}},\ \bibinfo {pages} {527--539} (\bibinfo {year}
  {1967})}\BibitemShut {NoStop}%
\bibitem [{\citenamefont {Schmeeckle}, \citenamefont {Nelson},\ and\
  \citenamefont {Shreve}(2007)}]{Schmeeckle2007}%
  \BibitemOpen
  \bibfield  {author} {\bibinfo {author} {\bibfnamefont {M.~W.}\ \bibnamefont
  {Schmeeckle}}, \bibinfo {author} {\bibfnamefont {J.~M.}\ \bibnamefont
  {Nelson}}, \ and\ \bibinfo {author} {\bibfnamefont {R.~L.}\ \bibnamefont
  {Shreve}},\ }\bibfield  {title} {\enquote {\bibinfo {title} {Forces on
  stationary particles in near-bed turbulent flows},}\ }\href {\doibase
  10.1029/2006JF000536} {\bibfield  {journal} {\bibinfo  {journal} {Journal of
  Geophysical Research: Earth Surface}\ }\textbf {\bibinfo {volume} {112}},\
  \bibinfo {pages} {F02003} (\bibinfo {year} {2007})}\BibitemShut {NoStop}%
\bibitem [{\citenamefont {Wiberg}\ and\ \citenamefont
  {Smith}(1985)}]{Wiberg1985}%
  \BibitemOpen
  \bibfield  {author} {\bibinfo {author} {\bibfnamefont {P.~L.}\ \bibnamefont
  {Wiberg}}\ and\ \bibinfo {author} {\bibfnamefont {J.~D.}\ \bibnamefont
  {Smith}},\ }\bibfield  {title} {\enquote {\bibinfo {title} {A theoretical
  model for saltating grains in water},}\ }\href {\doibase
  10.1029/JC090iC04p07341} {\bibfield  {journal} {\bibinfo  {journal} {Journal
  of Geophysical Research: Oceans}\ }\textbf {\bibinfo {volume} {90}},\
  \bibinfo {pages} {7341--7354} (\bibinfo {year} {1985})}\BibitemShut {NoStop}%
\bibitem [{\citenamefont {DallaValle}(1948)}]{Dallavalle1948}%
  \BibitemOpen
  \bibfield  {author} {\bibinfo {author} {\bibfnamefont {J.~M.}\ \bibnamefont
  {DallaValle}},\ }\href@noop {} {\emph {\bibinfo {title} {Micrometrics : The
  technology of fine particles}}},\ Vol.\ \bibinfo {volume} {2nd edition}\
  (\bibinfo  {publisher} {Pitman Pub. Corp},\ \bibinfo {year}
  {1948})\BibitemShut {NoStop}%
\bibitem [{\citenamefont {Richardson}\ and\ \citenamefont
  {Zaki}(1954)}]{Richardson1954}%
  \BibitemOpen
  \bibfield  {author} {\bibinfo {author} {\bibfnamefont {J.~F.}\ \bibnamefont
  {Richardson}}\ and\ \bibinfo {author} {\bibfnamefont {W.~N.}\ \bibnamefont
  {Zaki}},\ }\bibfield  {title} {\enquote {\bibinfo {title} {Sedimentation and
  fluidization: Part i},}\ }\href {\doibase 10.1016/S0263-8762(97)80006-8}
  {\bibfield  {journal} {\bibinfo  {journal} {Trans. Instn. Chem. Engrs}\
  }\textbf {\bibinfo {volume} {32}} (\bibinfo {year} {1954}),\
  10.1016/S0263-8762(97)80006-8}\BibitemShut {NoStop}%
\bibitem [{\citenamefont {Jackson}(1997)}]{Jackson1997}%
  \BibitemOpen
  \bibfield  {author} {\bibinfo {author} {\bibfnamefont {R.}~\bibnamefont
  {Jackson}},\ }\bibfield  {title} {\enquote {\bibinfo {title} {Locally
  averaged equations of motion for a mixture of identical spherical particles
  and a newtonian fluid},}\ }\href {\doibase 10.1016/S0009-2509(97)00065-1}
  {\bibfield  {journal} {\bibinfo  {journal} {Chemical Engineering Science}\
  }\textbf {\bibinfo {volume} {52}},\ \bibinfo {pages} {2457 -- 2469} (\bibinfo
  {year} {1997})}\BibitemShut {NoStop}%
\bibitem [{\citenamefont {Einstein}(1906)}]{Einstein1906}%
  \BibitemOpen
  \bibfield  {author} {\bibinfo {author} {\bibfnamefont {A.}~\bibnamefont
  {Einstein}},\ }\bibfield  {title} {\enquote {\bibinfo {title} {Eine neue
  bestimmung der molek\"ul dimensionen},}\ }\href@noop {} {\bibfield  {journal}
  {\bibinfo  {journal} {Ann. Physik}\ }\textbf {\bibinfo {volume} {19}},\
  \bibinfo {pages} {289--306} (\bibinfo {year} {1906})}\BibitemShut {NoStop}%
\bibitem [{\citenamefont {Zannetti}(1986)}]{Zannetti1986}%
  \BibitemOpen
  \bibfield  {author} {\bibinfo {author} {\bibfnamefont {P.}~\bibnamefont
  {Zannetti}},\ }\bibfield  {title} {\enquote {\bibinfo {title} {Monte-carlo
  simulation of auto- and cross-correlated turbulent velocity fluctuations
  (mc-lagpar \{II\} model)},}\ }\href {\doibase 10.1016/0266-9838(86)90033-X}
  {\bibfield  {journal} {\bibinfo  {journal} {Environmental Software}\ }\textbf
  {\bibinfo {volume} {1}},\ \bibinfo {pages} {26 -- 30} (\bibinfo {year}
  {1986})}\BibitemShut {NoStop}%
\bibitem [{\citenamefont {Nezu}(1977)}]{Nezu1977}%
  \BibitemOpen
  \bibfield  {author} {\bibinfo {author} {\bibfnamefont {I.}~\bibnamefont
  {Nezu}},\ }\emph {\bibinfo {title} {Turbulent structure in open-channel
  flows}},\ \href
  {http://repository.tudelft.nl/view/hydro/uuid%3Aa41f39c2-fce6-4647-bd7a-1d412c720ed7/}
  {Ph.D. thesis},\ \bibinfo  {school} {Kyoto University} (\bibinfo {year}
  {1977})\BibitemShut {NoStop}%
\bibitem [{\citenamefont {Nezu}\ and\ \citenamefont
  {Nakagawa}(1993)}]{Nezu1993}%
  \BibitemOpen
  \bibfield  {author} {\bibinfo {author} {\bibfnamefont {I.}~\bibnamefont
  {Nezu}}\ and\ \bibinfo {author} {\bibfnamefont {H.}~\bibnamefont
  {Nakagawa}},\ }\href@noop {} {\emph {\bibinfo {title} {Turbulence in Open
  Channel Flows}}}\ (\bibinfo  {publisher} {Taylor \& Francis},\ \bibinfo
  {year} {1993})\BibitemShut {NoStop}%
\bibitem [{\citenamefont {Chauchat}\ \emph {et~al.}(2013)\citenamefont
  {Chauchat}, \citenamefont {Guillou}, \citenamefont {Pham Van~Bang},\ and\
  \citenamefont {Dan~Nguyen}}]{Chauchat2013}%
  \BibitemOpen
  \bibfield  {author} {\bibinfo {author} {\bibfnamefont {J.}~\bibnamefont
  {Chauchat}}, \bibinfo {author} {\bibfnamefont {S.}~\bibnamefont {Guillou}},
  \bibinfo {author} {\bibfnamefont {D.}~\bibnamefont {Pham Van~Bang}}, \ and\
  \bibinfo {author} {\bibfnamefont {K.}~\bibnamefont {Dan~Nguyen}},\ }\bibfield
   {title} {\enquote {\bibinfo {title} {Modelling sedimentation-consolidation
  in the framework of a one-dimensional two-phase flow model},}\ }\href
  {\doibase 10.1080/00221686.2013.768798} {\bibfield  {journal} {\bibinfo
  {journal} {Journal of Hydraulic Research}\ }\textbf {\bibinfo {volume}
  {51}},\ \bibinfo {pages} {293--305} (\bibinfo {year} {2013})}\BibitemShut
  {NoStop}%
\bibitem [{\citenamefont {\v{S}milauer}\ \emph {et~al.}(2010)\citenamefont
  {\v{S}milauer}, \citenamefont {Catalano}, \citenamefont {Chareyre},
  \citenamefont {Dorofeenko}, \citenamefont {Duriez}, \citenamefont {Gladky},
  \citenamefont {Kozicki}, \citenamefont {Modenese}, \citenamefont
  {Scholt\`es}, \citenamefont {Sibille}, \citenamefont {Str\'ansk\'y},\ and\
  \citenamefont {Thoeni}}]{YADE2010}%
  \BibitemOpen
  \bibfield  {author} {\bibinfo {author} {\bibfnamefont {V.}~\bibnamefont
  {\v{S}milauer}}, \bibinfo {author} {\bibfnamefont {E.}~\bibnamefont
  {Catalano}}, \bibinfo {author} {\bibfnamefont {B.}~\bibnamefont {Chareyre}},
  \bibinfo {author} {\bibfnamefont {S.}~\bibnamefont {Dorofeenko}}, \bibinfo
  {author} {\bibfnamefont {J.}~\bibnamefont {Duriez}}, \bibinfo {author}
  {\bibfnamefont {A.}~\bibnamefont {Gladky}}, \bibinfo {author} {\bibfnamefont
  {J.}~\bibnamefont {Kozicki}}, \bibinfo {author} {\bibfnamefont
  {C.}~\bibnamefont {Modenese}}, \bibinfo {author} {\bibfnamefont
  {L.}~\bibnamefont {Scholt\`es}}, \bibinfo {author} {\bibfnamefont
  {L.}~\bibnamefont {Sibille}}, \bibinfo {author} {\bibfnamefont
  {J.}~\bibnamefont {Str\'ansk\'y}}, \ and\ \bibinfo {author} {\bibfnamefont
  {K.}~\bibnamefont {Thoeni}},\ }\href {http://yade-dem.org/doc/} {\emph
  {\bibinfo {title} {Yade Documentation (V. \v{S}milauer, ed.), The Yade
  Project, 1st ed.}}} (\bibinfo {year} {2010})\BibitemShut {NoStop}%
\bibitem [{\citenamefont {Bathe}\ and\ \citenamefont
  {Wilson}(1976)}]{Bathe1976}%
  \BibitemOpen
  \bibfield  {author} {\bibinfo {author} {\bibfnamefont {K.-J.}\ \bibnamefont
  {Bathe}}\ and\ \bibinfo {author} {\bibfnamefont {E.}~\bibnamefont {Wilson}},\
  }\href@noop {} {\emph {\bibinfo {title} {Numerical methods in finite element
  analysis}}}\ (\bibinfo  {publisher} {Prentice-Hall (Englewood Cliffs,
  N.J.)},\ \bibinfo {year} {1976})\BibitemShut {NoStop}%
\bibitem [{\citenamefont {Catalano}(2012)}]{Catalano2012}%
  \BibitemOpen
  \bibfield  {author} {\bibinfo {author} {\bibfnamefont {E.}~\bibnamefont
  {Catalano}},\ }\emph {\bibinfo {title} {A pore-scale coupled hydromechanical
  model for biphasic granular media. Application to granular sediment
  hydrodynamics}},\ \href {https://yade-dem.org/publi/Catalano_Thesis.pdf}
  {Ph.D. thesis},\ \bibinfo  {school} {Universit\'e de Grenoble} (\bibinfo
  {year} {2012})\BibitemShut {NoStop}%
\bibitem [{\citenamefont {Catalano}, \citenamefont {Chareyre},\ and\
  \citenamefont {Barth{\'e}l{\'e}my}(2014)}]{Catalano2014}%
  \BibitemOpen
  \bibfield  {author} {\bibinfo {author} {\bibfnamefont {E.}~\bibnamefont
  {Catalano}}, \bibinfo {author} {\bibfnamefont {B.}~\bibnamefont {Chareyre}},
  \ and\ \bibinfo {author} {\bibfnamefont {E.}~\bibnamefont
  {Barth{\'e}l{\'e}my}},\ }\bibfield  {title} {\enquote {\bibinfo {title}
  {Pore-scale modeling of fluid-particles interaction and emerging
  poromechanical effects},}\ }\href {\doibase 10.1002/nag.2198} {\bibfield
  {journal} {\bibinfo  {journal} {Int. Journal for Numerical and Analytical
  Methods in Geomechanics}\ }\textbf {\bibinfo {volume} {38}},\ \bibinfo
  {pages} {51--71} (\bibinfo {year} {2014})}\BibitemShut {NoStop}%
\bibitem [{\citenamefont {Chareyre}\ and\ \citenamefont
  {Villard}(2005)}]{Chareyre2005}%
  \BibitemOpen
  \bibfield  {author} {\bibinfo {author} {\bibfnamefont {B.}~\bibnamefont
  {Chareyre}}\ and\ \bibinfo {author} {\bibfnamefont {P.}~\bibnamefont
  {Villard}},\ }\bibfield  {title} {\enquote {\bibinfo {title} {Dynamic spar
  elements and discrete element methods in two dimensions for the modeling of
  soil-inclusion problems},}\ }\href {\doibase
  10.1061/(ASCE)0733-9399(2005)131:7(689)} {\bibfield  {journal} {\bibinfo
  {journal} {Journal of engineering mechanics}\ }\textbf {\bibinfo {volume}
  {131}},\ \bibinfo {pages} {689} (\bibinfo {year} {2005})}\BibitemShut
  {NoStop}%
\bibitem [{\citenamefont {B\"ohm}\ \emph {et~al.}(2006)\citenamefont {B\"ohm},
  \citenamefont {Frey}, \citenamefont {Ducottet}, \citenamefont {Ancey},
  \citenamefont {Jodeau},\ and\ \citenamefont {Reboud}}]{Bohm2006}%
  \BibitemOpen
  \bibfield  {author} {\bibinfo {author} {\bibfnamefont {T.}~\bibnamefont
  {B\"ohm}}, \bibinfo {author} {\bibfnamefont {P.}~\bibnamefont {Frey}},
  \bibinfo {author} {\bibfnamefont {C.}~\bibnamefont {Ducottet}}, \bibinfo
  {author} {\bibfnamefont {C.}~\bibnamefont {Ancey}}, \bibinfo {author}
  {\bibfnamefont {M.}~\bibnamefont {Jodeau}}, \ and\ \bibinfo {author}
  {\bibfnamefont {J.}~\bibnamefont {Reboud}},\ }\bibfield  {title} {\enquote
  {\bibinfo {title} {Two-dimensional motion of a set of particles in a free
  surface flow with image processing},}\ }\href {\doibase
  10.1007/s00348-006-0134-9} {\bibfield  {journal} {\bibinfo  {journal}
  {Experiments in Fluids}\ }\textbf {\bibinfo {volume} {41}},\ \bibinfo {pages}
  {1--11} (\bibinfo {year} {2006})}\BibitemShut {NoStop}%
\bibitem [{\citenamefont {Hergault}\ \emph {et~al.}(2010)\citenamefont
  {Hergault}, \citenamefont {Frey}, \citenamefont {M\'etivier}, \citenamefont
  {Barat}, \citenamefont {Ducottet}, \citenamefont {B\"ohm},\ and\
  \citenamefont {Ancey}}]{Hergault2010}%
  \BibitemOpen
  \bibfield  {author} {\bibinfo {author} {\bibfnamefont {V.}~\bibnamefont
  {Hergault}}, \bibinfo {author} {\bibfnamefont {P.}~\bibnamefont {Frey}},
  \bibinfo {author} {\bibfnamefont {F.}~\bibnamefont {M\'etivier}}, \bibinfo
  {author} {\bibfnamefont {C.}~\bibnamefont {Barat}}, \bibinfo {author}
  {\bibfnamefont {C.}~\bibnamefont {Ducottet}}, \bibinfo {author}
  {\bibfnamefont {T.}~\bibnamefont {B\"ohm}}, \ and\ \bibinfo {author}
  {\bibfnamefont {C.}~\bibnamefont {Ancey}},\ }\bibfield  {title} {\enquote
  {\bibinfo {title} {Image processing for the study of bedload transport of
  two-size spherical particles in a supercritical flow},}\ }\href {\doibase
  10.1007/s00348-010-0856-6} {\bibfield  {journal} {\bibinfo  {journal}
  {Experiments in Fluids}\ }\textbf {\bibinfo {volume} {49}},\ \bibinfo {pages}
  {1095--1107} (\bibinfo {year} {2010})}\BibitemShut {NoStop}%
\bibitem [{\citenamefont {Frey}\ and\ \citenamefont {Reboud}(2001)}]{Frey2001}%
  \BibitemOpen
  \bibfield  {author} {\bibinfo {author} {\bibfnamefont {P.}~\bibnamefont
  {Frey}}\ and\ \bibinfo {author} {\bibfnamefont {J.-L.}\ \bibnamefont
  {Reboud}},\ }\bibfield  {title} {\enquote {\bibinfo {title} {Experimental
  study of narrow free-surface turbulent flows on steep slopes},}\ }in\
  \href@noop {} {\emph {\bibinfo {booktitle} {Advances in flow modeling and
  turbulence measurements}}},\ \bibinfo {editor} {edited by\ \bibinfo {editor}
  {\bibfnamefont {T.~N.}\ \bibnamefont {Ninokata~H}, \bibfnamefont {Wada~A}}}\
  (\bibinfo  {publisher} {World Scientific Publishing Co.},\ \bibinfo {address}
  {Tokyo},\ \bibinfo {year} {2001})\ pp.\ \bibinfo {pages}
  {396--403}\BibitemShut {NoStop}%
\bibitem [{\citenamefont {Graf}\ and\ \citenamefont
  {Altinakar}(1998)}]{Graf1998}%
  \BibitemOpen
  \bibfield  {author} {\bibinfo {author} {\bibfnamefont {W.}~\bibnamefont
  {Graf}}\ and\ \bibinfo {author} {\bibfnamefont {S.}~\bibnamefont
  {Altinakar}},\ }\href@noop {} {\emph {\bibinfo {title} {Fluvial hydraulics:
  flow and transport processes in channels of simple geometry}}}\ (\bibinfo
  {publisher} {Wiley},\ \bibinfo {year} {1998})\BibitemShut {NoStop}%
\bibitem [{\citenamefont {Frey}\ \emph {et~al.}(2006)\citenamefont {Frey},
  \citenamefont {Dufresne}, \citenamefont {B\"ohm}, \citenamefont {Jodeau},\
  and\ \citenamefont {Ancey}}]{Frey2006}%
  \BibitemOpen
  \bibfield  {author} {\bibinfo {author} {\bibfnamefont {P.}~\bibnamefont
  {Frey}}, \bibinfo {author} {\bibfnamefont {M.}~\bibnamefont {Dufresne}},
  \bibinfo {author} {\bibfnamefont {T.}~\bibnamefont {B\"ohm}}, \bibinfo
  {author} {\bibfnamefont {M.}~\bibnamefont {Jodeau}}, \ and\ \bibinfo {author}
  {\bibfnamefont {C.}~\bibnamefont {Ancey}},\ }\bibfield  {title} {\enquote
  {\bibinfo {title} {Experimental study of bed load on steep slopes},}\ }in\
  \href@noop {} {\emph {\bibinfo {booktitle} {River Flow}}},\ \bibinfo {editor}
  {edited by\ \bibinfo {editor} {\bibnamefont {Ferreira}}, \bibinfo {editor}
  {\bibnamefont {Alves}}, \bibinfo {editor} {\bibnamefont {Leal}}, \ and\
  \bibinfo {editor} {\bibnamefont {Cardoso}}}\ (\bibinfo  {publisher} {Taylor
  \& Francis Group},\ \bibinfo {address} {Lisbonne, Portugal},\ \bibinfo {year}
  {2006})\ pp.\ \bibinfo {pages} {887--893}\BibitemShut {NoStop}%
\bibitem [{\citenamefont {Bigillon}(2001)}]{Bigillon2001}%
  \BibitemOpen
  \bibfield  {author} {\bibinfo {author} {\bibfnamefont {F.}~\bibnamefont
  {Bigillon}},\ }\emph {\bibinfo {title} {Etude du mouvement bidimensionnel
  d'une particule dans un courant d'eau sur forte pente}},\ \href@noop {}
  {Ph.D. thesis},\ \bibinfo  {school} {Universit\'e Joseph Fourier} (\bibinfo
  {year} {2001})\BibitemShut {NoStop}%
\bibitem [{\citenamefont {Roux}\ and\ \citenamefont {Combe}(2002)}]{Roux2002}%
  \BibitemOpen
  \bibfield  {author} {\bibinfo {author} {\bibfnamefont {J.-N.}\ \bibnamefont
  {Roux}}\ and\ \bibinfo {author} {\bibfnamefont {G.}~\bibnamefont {Combe}},\
  }\bibfield  {title} {\enquote {\bibinfo {title} {Quasistatic rheology and the
  origins of strain},}\ }\href {\doibase 10.1016/S1631-0705(02)01306-3}
  {\bibfield  {journal} {\bibinfo  {journal} {Comptes Rendus Physique}\
  }\textbf {\bibinfo {volume} {3}},\ \bibinfo {pages} {131 -- 140} (\bibinfo
  {year} {2002})}\BibitemShut {NoStop}%
\bibitem [{\citenamefont {Lajeunesse}, \citenamefont {Malverti},\ and\
  \citenamefont {Charru}(2010)}]{Lajeunesse2010}%
  \BibitemOpen
  \bibfield  {author} {\bibinfo {author} {\bibfnamefont {E.}~\bibnamefont
  {Lajeunesse}}, \bibinfo {author} {\bibfnamefont {L.}~\bibnamefont
  {Malverti}}, \ and\ \bibinfo {author} {\bibfnamefont {F.}~\bibnamefont
  {Charru}},\ }\bibfield  {title} {\enquote {\bibinfo {title} {Bed load
  transport in turbulent flow at the grain scale: Experiments and modeling},}\
  }\href {\doibase 10.1029/2009JF001628} {\bibfield  {journal} {\bibinfo
  {journal} {Journal of Geophysical Research: Earth Surface}\ }\textbf
  {\bibinfo {volume} {115}},\ \bibinfo {pages} {F4} (\bibinfo {year}
  {2010})}\BibitemShut {NoStop}%
\bibitem [{\citenamefont {Wilson}(1966)}]{Wilson1966}%
  \BibitemOpen
  \bibfield  {author} {\bibinfo {author} {\bibfnamefont {K.~C.}\ \bibnamefont
  {Wilson}},\ }\bibfield  {title} {\enquote {\bibinfo {title} {Bed-load
  transport at high shear stress},}\ }in\ \href@noop {} {\emph {\bibinfo
  {booktitle} {A.S.C.E}}},\ Vol.\ \bibinfo {volume} {HY6},\ \bibinfo {editor}
  {edited by\ \bibinfo {editor} {\bibnamefont {ASCE}}}\ (\bibinfo {year}
  {1966})\BibitemShut {NoStop}%
\bibitem [{\citenamefont {Yalin}(1977)}]{Yalin1977}%
  \BibitemOpen
  \bibfield  {author} {\bibinfo {author} {\bibfnamefont {M.~S.}\ \bibnamefont
  {Yalin}},\ }\href@noop {} {\emph {\bibinfo {title} {Mechanics of sediment
  transport}}},\ \bibinfo {edition} {2nd}\ ed.\ (\bibinfo  {publisher}
  {Pergamon Press},\ \bibinfo {address} {Ontario},\ \bibinfo {year}
  {1977})\BibitemShut {NoStop}%
\bibitem [{\citenamefont {Wilson}(1987)}]{Wilson1987}%
  \BibitemOpen
  \bibfield  {author} {\bibinfo {author} {\bibfnamefont {K.}~\bibnamefont
  {Wilson}},\ }\bibfield  {title} {\enquote {\bibinfo {title} {Analysis of
  bed‐load motion at high shear stress},}\ }\href {\doibase
  10.1061/(ASCE)0733-9429(1987)113:1(97)} {\bibfield  {journal} {\bibinfo
  {journal} {Journal of Hydraulic Engineering}\ }\textbf {\bibinfo {volume}
  {113}},\ \bibinfo {pages} {97--103} (\bibinfo {year} {1987})}\BibitemShut
  {NoStop}%
\bibitem [{\citenamefont {Buffington}\ and\ \citenamefont
  {Montgomery}(1997)}]{Buffington1997}%
  \BibitemOpen
  \bibfield  {author} {\bibinfo {author} {\bibfnamefont {J.}~\bibnamefont
  {Buffington}}\ and\ \bibinfo {author} {\bibfnamefont {D.}~\bibnamefont
  {Montgomery}},\ }\bibfield  {title} {\enquote {\bibinfo {title} {A systematic
  analysis of eight decades of incipient motion studies, with special reference
  to gravel-bedded rivers},}\ }\href {\doibase 10.1029/96WR03190} {\bibfield
  {journal} {\bibinfo  {journal} {Water Resources Research}\ }\textbf {\bibinfo
  {volume} {33}},\ \bibinfo {pages} {1993--2029} (\bibinfo {year}
  {1997})}\BibitemShut {NoStop}%
\bibitem [{\citenamefont {Ouriemi}\ \emph {et~al.}(2007)\citenamefont
  {Ouriemi}, \citenamefont {Aussillous}, \citenamefont {Medale}, \citenamefont
  {Peysson},\ and\ \citenamefont {Guazzelli}}]{Ouriemi2007}%
  \BibitemOpen
  \bibfield  {author} {\bibinfo {author} {\bibfnamefont {M.}~\bibnamefont
  {Ouriemi}}, \bibinfo {author} {\bibfnamefont {P.}~\bibnamefont {Aussillous}},
  \bibinfo {author} {\bibfnamefont {M.}~\bibnamefont {Medale}}, \bibinfo
  {author} {\bibfnamefont {Y.}~\bibnamefont {Peysson}}, \ and\ \bibinfo
  {author} {\bibfnamefont {Ã.}~\bibnamefont {Guazzelli}},\ }\bibfield  {title}
  {\enquote {\bibinfo {title} {Determination of the critical shields number for
  particle erosion in laminar flow},}\ }\href {\doibase 10.1063/1.2747677}
  {\bibfield  {journal} {\bibinfo  {journal} {Physics of Fluids}\ }\textbf
  {\bibinfo {volume} {19}},\ \bibinfo {eid} {061706} (\bibinfo {year}
  {2007})}\BibitemShut {NoStop}%
\bibitem [{\citenamefont {Papanicolaou}\ \emph {et~al.}(2002)\citenamefont
  {Papanicolaou}, \citenamefont {Diplas}, \citenamefont {Evaggelopoulos},\ and\
  \citenamefont {Fotopoulos}}]{Papanicolaou2002}%
  \BibitemOpen
  \bibfield  {author} {\bibinfo {author} {\bibfnamefont {A.}~\bibnamefont
  {Papanicolaou}}, \bibinfo {author} {\bibfnamefont {P.}~\bibnamefont
  {Diplas}}, \bibinfo {author} {\bibfnamefont {N.}~\bibnamefont
  {Evaggelopoulos}}, \ and\ \bibinfo {author} {\bibfnamefont {S.}~\bibnamefont
  {Fotopoulos}},\ }\bibfield  {title} {\enquote {\bibinfo {title} {Stochastic
  incipient motion criterion for spheres under various bed packing
  conditions},}\ }\href {\doibase 10.1061/(ASCE)0733-9429(2002)128:4(369)}
  {\bibfield  {journal} {\bibinfo  {journal} {Journal of Hydraulic
  Engineering}\ }\textbf {\bibinfo {volume} {128}},\ \bibinfo {pages}
  {369--380} (\bibinfo {year} {2002})}\BibitemShut {NoStop}%
\bibitem [{\citenamefont {Dwivedi}\ \emph {et~al.}(2012)\citenamefont
  {Dwivedi}, \citenamefont {Melville}, \citenamefont {Raudkivi}, \citenamefont
  {Shamseldin},\ and\ \citenamefont {Chiew}}]{Dwivedi2012}%
  \BibitemOpen
  \bibfield  {author} {\bibinfo {author} {\bibfnamefont {A.}~\bibnamefont
  {Dwivedi}}, \bibinfo {author} {\bibfnamefont {B.}~\bibnamefont {Melville}},
  \bibinfo {author} {\bibfnamefont {A.}~\bibnamefont {Raudkivi}}, \bibinfo
  {author} {\bibfnamefont {A.}~\bibnamefont {Shamseldin}}, \ and\ \bibinfo
  {author} {\bibfnamefont {Y.}~\bibnamefont {Chiew}},\ }\bibfield  {title}
  {\enquote {\bibinfo {title} {Role of turbulence and particle exposure on
  entrainment of large spherical particles in flows with low relative
  submergence},}\ }\href {\doibase 10.1061/(ASCE)HY.1943-7900.0000632}
  {\bibfield  {journal} {\bibinfo  {journal} {Journal of Hydraulic
  Engineering}\ }\textbf {\bibinfo {volume} {138}},\ \bibinfo {pages}
  {1022--1030} (\bibinfo {year} {2012})}\BibitemShut {NoStop}%
\bibitem [{\citenamefont {Gondret}, \citenamefont {Lance},\ and\ \citenamefont
  {Petit}(2002)}]{Gondret2002}%
  \BibitemOpen
  \bibfield  {author} {\bibinfo {author} {\bibfnamefont {P.}~\bibnamefont
  {Gondret}}, \bibinfo {author} {\bibfnamefont {M.}~\bibnamefont {Lance}}, \
  and\ \bibinfo {author} {\bibfnamefont {L.}~\bibnamefont {Petit}},\ }\bibfield
   {title} {\enquote {\bibinfo {title} {Bouncing motion of spherical particles
  in fluids},}\ }\href {\doibase {10.1063/1.1427920}} {\bibfield  {journal}
  {\bibinfo  {journal} {Physics of Fluids}\ }\textbf {\bibinfo {volume} {14}},\
  \bibinfo {pages} {643--652} (\bibinfo {year} {2002})}\BibitemShut {NoStop}%
\bibitem [{\citenamefont {Armanini}\ \emph {et~al.}(2005)\citenamefont
  {Armanini}, \citenamefont {Capart}, \citenamefont {Fraccarollo},\ and\
  \citenamefont {Larcher}}]{Armanini2005}%
  \BibitemOpen
  \bibfield  {author} {\bibinfo {author} {\bibfnamefont {A.}~\bibnamefont
  {Armanini}}, \bibinfo {author} {\bibfnamefont {H.}~\bibnamefont {Capart}},
  \bibinfo {author} {\bibfnamefont {L.}~\bibnamefont {Fraccarollo}}, \ and\
  \bibinfo {author} {\bibfnamefont {M.}~\bibnamefont {Larcher}},\ }\bibfield
  {title} {\enquote {\bibinfo {title} {Rheological stratification in
  experimental free-surface flows of granular liquid mixtures},}\ }\href
  {\doibase 10.1017/S0022112005004283} {\bibfield  {journal} {\bibinfo
  {journal} {Journal of Fluid Mechanics}\ }\textbf {\bibinfo {volume} {532}},\
  \bibinfo {pages} {269--319} (\bibinfo {year} {2005})}\BibitemShut {NoStop}%
\bibitem [{\citenamefont {da~Cruz}\ \emph {et~al.}(2005)\citenamefont
  {da~Cruz}, \citenamefont {Emam}, \citenamefont {Prochnow}, \citenamefont
  {Roux},\ and\ \citenamefont {Chevoir}}]{DaCruz2005}%
  \BibitemOpen
  \bibfield  {author} {\bibinfo {author} {\bibfnamefont {F.}~\bibnamefont
  {da~Cruz}}, \bibinfo {author} {\bibfnamefont {S.}~\bibnamefont {Emam}},
  \bibinfo {author} {\bibfnamefont {M.}~\bibnamefont {Prochnow}}, \bibinfo
  {author} {\bibfnamefont {J.-N.}\ \bibnamefont {Roux}}, \ and\ \bibinfo
  {author} {\bibfnamefont {F.}~\bibnamefont {Chevoir}},\ }\bibfield  {title}
  {\enquote {\bibinfo {title} {Rheophysics of dense granular materials:
  Discrete simulation of plane shear flows},}\ }\href {\doibase
  10.1103/PhysRevE.72.021309} {\bibfield  {journal} {\bibinfo  {journal} {Phys.
  Rev. E}\ }\textbf {\bibinfo {volume} {72}},\ \bibinfo {pages} {021309}
  (\bibinfo {year} {2005})}\BibitemShut {NoStop}%
\bibitem [{\citenamefont {Hill}, \citenamefont {Gioia},\ and\ \citenamefont
  {Tota}(2003)}]{Hill2003}%
  \BibitemOpen
  \bibfield  {author} {\bibinfo {author} {\bibfnamefont {K.~M.}\ \bibnamefont
  {Hill}}, \bibinfo {author} {\bibfnamefont {G.}~\bibnamefont {Gioia}}, \ and\
  \bibinfo {author} {\bibfnamefont {V.~V.}\ \bibnamefont {Tota}},\ }\bibfield
  {title} {\enquote {\bibinfo {title} {Structure and kinematics in dense
  free-surface granular flow},}\ }\href {\doibase
  10.1103/PhysRevLett.91.064302} {\bibfield  {journal} {\bibinfo  {journal}
  {Phys. Rev. Lett.}\ }\textbf {\bibinfo {volume} {91}},\ \bibinfo {pages}
  {064302} (\bibinfo {year} {2003})}\BibitemShut {NoStop}%
\end{thebibliography}%

\end{document}